\documentclass[12pt, draftclsnofoot, onecolumn]{IEEEtran}
\usepackage[cmex10]{amsmath}
\usepackage{array}
\usepackage{mdwmath}
\usepackage{mdwtab}
\usepackage{eqparbox}

\usepackage{bm}
\usepackage{multicol}
\usepackage{multirow}
\usepackage{hhline} 

\usepackage{color, soul}
\usepackage{framed}

\newcommand{\FB}[1]{\textcolor{black}{#1}} 
\newcommand{\FBB}[1]{\textcolor{black}{#1}} 

\usepackage{times,amsmath,epsfig}
\usepackage{amsmath}
\usepackage{amsthm}
\usepackage{amscd}
\usepackage{amssymb}
\usepackage{graphicx}
\usepackage[tight,footnotesize]{subfigure}
\usepackage{epstopdf}

\usepackage{rotating}

\usepackage{hyperref}
\usepackage{tablefootnote}
\usepackage{tabularx, booktabs}
\usepackage{cancel}
\usepackage[normalem]{ulem}

\usepackage[numbers,sort&compress]{natbib}

\ifCLASSINFOpdf
\else
\fi
\begin{document}
%
\title{\huge Doppler Spectrum Analysis of a Roadside Scatterer Model for Vehicle-to-Vehicle Channels: An Indirect Method}

\author{Sangjo Yoo, David Gonz\'alez G., Jyri H\"{a}m\"{a}l\"{a}inen, and~Kiseon Kim

\thanks{S. Yoo and K. Kim are with the School of Electrical Engineering and Computer Science, Gwangju Institute of Science and Technology (GIST), Gwangju, 61005, Republic of Korea. Email:\{asapyoo, kskim\}@gist.ac.kr.}
\thanks{D. Gonz\'alez G. and J. H\"{a}m\"{a}l\"{a}inen are with the Department of Communications and Networking, Aalto University, Espoo, 00076, Finland. Email: david.gonzalez.g@ieee.org, jyri.hamalainen@aalto.fi.}
\thanks{This work has been submitted to the IEEE for possible publication. Copyright may be transferred without notice, after which this version may no longer be accessible.}
       
%

}
\markboth{}%
{Shell \MakeLowercase{\textit{et al.}}: Bare Demo of IEEEtran.cls for Journals}



%


\maketitle

\begin{abstract}
In vehicle-to-vehicle (V2V) channels, such roadside scatterers (RSSs) as houses, buildings, trees, and many more, play a crucial role in the determination of the Doppler power spectral density (DPSD) characteristics. However, the relevant research results are scarce due to the lack of computationally tractable analytic DPSD solutions. To fill this gap, we investigate an indirect method for the DPSD analysis of a generic two-dimensional (2D) RSS model for V2V channels. The indirect method, based on the Hoeher's theorem, employs successive transformations of random variables to obtain the DPSD. \FBB{Compared to the conventional methods, leading to impractical multiple integral solutions, our method yields a single integral-form, more useful for analytic studies, model validation/parameter estimation, and fading simulator design.} 
Using the new DPSD solution, the impact of different RSS layouts on the DPSD characteristics is further investigated, and several new insights are provided. The joint probability density function (PDF) of angle-of-departure and angle-of-arrival (AoA) and the joint Doppler-AoA PDF are newly presented in closed-forms and analyzed with respect to the DPSD shape. Comparisons with the DPSDs measured in highway and \FB{urban canyon} environments demonstrate not only the validity of the generic 2D RSS model, but also the significant contribution of RSSs to V2V channels. 
\end{abstract}


%
\IEEEpeerreviewmaketitle

\section{Introduction}

\FBB{In the literature, Doppler power spectral density (DPSD) analysis for vehicle-to-vehicle (V2V) channels has been performed either based on real measurements \cite{Aco07, Aco07_2, Tan08, Zaj09_M, Che13} or analytical derivations from geometry-based stochastic channel models (GBSCMs) \cite{Akk86, Zaj15, Zaj09,  Che09_2, Zaj14, XChe13, Pat05, Zaj08, Yua14, Zha16, Patbook11, Zho12, Ava12, Ava11}. The former provides a ground truth while the latter case (that is of our interest) provides an analytic way to investigate how dynamics of the transmitter (Tx) and receiver (Rx), as well as  scatterer geometries can impact on V2V channels in the Doppler frequency domain, in relation to the physical and geometrical model parameters. 
On the other hand, analytic DPSD solutions of such GBSCMs are used in numerical optimizations for model validation (or parameter estimation) using measurements \cite{Zaj09, Che09_2, Zaj14, XChe13} and fading simulator development (e.g. Doppler filter design \cite{Ali12}), which are important prerequisite for realistic, yet efficient V2V system simulations \cite{Patbook11}. Hence, finding accurate and tractable analytic DPSD solutions of GBSCMs, reflecting realistic V2V environments, is an important research problem in both theoretical and practical aspects.}

%
\FBB{There are some number of works already done for DPSD analysis in V2V channels. However, most of the previous works are based on the channel models using regular geometries}, such as two-rings, ellipses, two-cylinders, two-spheres, \FB {and their combination \cite{Akk86, Pat05, Zaj08, Che09_2, Zaj14, Zaj15, XChe13, Zaj09, Yua14, Zha16}}. These models, classified as regular-shaped GBSCMs (RS-GBSCMs), are useful for DPSD analysis due to the dimension reduction for the scatterer location representation, as well as geometrical approximations \cite{Yoo16, Yoo16_2}, which lead to simple analytic solutions. 
Yet, placing all scatterers on the regular shapes does not capture many features of real-world scenarios \cite{Kar09}. In practice, moving scatterers (cars) exist on the road while stationary scatterers (e.g. houses, buildings, trees, and sound blockers) are rather distributed along the roadsides. In particular, the locations of such stationary roadside scatterers (RSSs), relative to the Tx and Rx positions, can be significantly changed, according to the road layout (width and length) and road types (straight road, T-junction, cross-junction, tunnel, and etc.). 
Hence, irregular-shaped GBSCMs (IS-GBSCMs), considering realistic road geometry and placement of RSSs as in \cite{Kar09, Czi10, Che13, Zho12, Che07, Ava12, Wal14, Ava16, The13, Ava11}, are more reasonable than the RS-GBSCMs\footnote{For example, in \cite{Lia16}, DPSD analysis was carried out based on a RS-GBSCM using an ellipse geometry under a uniform single modal angle-of-arrival (AoA) assumption. However, such a scatterer representation is over-simplistic to characterize the signal dispersion by RSSs in reality (see Figs. 7-10 of \cite{Che13} and Fig. \ref{hist_pdf_comp} in this paper). The ellipse model in general produces skewed U-shape DPSDs as in \cite{Che09_2}, which do not match with the measured spectral shapes, generally observed in straight road environments, see \cite{Che13,Tan08, Aco07, Zaj09_M, Aco07_2}.}. 
However, it is in general difficult to obtain analytic and computationally efficient DPSD solutions of the IS-GBSCMs due to \FBB{the larger number of random variables used to describe RSS locations.}

\FBB{The aim of this paper is to investigate the DPSD characteristics of V2V channels due to roadside scatterers (RSSs) for a straight road, which is the most elementary, yet important V2V scenario \cite{Kar09, Che13, Ber14}. To date, only handful results have been reported on this problem due to the complexities of the channel models and the corresponding DPSD solutions.}
The study in \cite{Kar09} has shown that placing stationary scatterers on a line parallel to the road can produce a joint delay-Doppler support, similar to the measurement data. Based on this observation, the authors proposed a two-dimensional (2D) RSS model, where RSSs are uniformly distributed within two symmetric rectangles on roadsides. Yet, the model's \FBB{analytic} DPSD was not investigated. Later, in \cite{Zho12}, the same model was used to analyze its DPSD through a simulation approach. However, the approach requires extensive simulations to obtain statistically reliable results. \FBB{Also, the estimated spectrum suffers from spectral leakage. Hence, the approach is impractical for the model validation/parameter estimation, and fading simulator developments, which require repetitive, accurate computations of the DPSD for different model parameter sets.}

To alleviate the issues posed by the simulation approach, a few analytic approaches were proposed based on one-dimensional (1D) \cite{Che13} and 2D RSS models \cite{Wal14, Ava12, Ava11}. In \cite{Che13}, 
an analytic DPSD solution was derived based on a radar equation, under unrealistic assumption that RSSs are placed on two infinite lines parallel to a road. In \cite{Ava12}, the authors formulated a channel gain function of the model and derived its analytic DPSD in a triple integral-form. Meanwhile, the authors in \cite{Ava11} derived the model's DPSD in a double integral-form by inverse Fourier transforming the product of the Tx and Rx Doppler frequency characteristic functions. Finally, in \cite{Wal14}, an algorithm was proposed for the computation of the delay-dependent Doppler frequency PDF (DPDF) under the assumption that RSSs are uniformly distributed on an equi-delay ellipse. 

It is noteworthy that the analytic approaches in \cite{Ava12, Wal14, Ava11} can be classified into two categories: the direct method in \cite{Ava12} and indirect method in \cite{Wal14, Ava11}. The former finds the DPSD of a channel gain function by Fourier transforming its auto-correlation function (ACF) under the Wiener-Khinchin theorem. The latter method alternatively find the DPDF based on the well-known proportionality between the DPSD and DPDF (i.e, Hoeher's theorem, see \cite{Hoe92}). \FBB{The direct method is a standard approach for DPSD analysis. However, it produces complex DPSD solutions for such 2D RSS models due to large number of random variables to be averaged via integral (statistical averaging) operations. In addition, the direct method requires a Fourier transform operation, involving an improper integral. The solution in \cite{Ava12}, obtained via the direct method, is indeed complex so even cannot be computed by conventional numerical integral solvers.}

\FBB{On the other hand, the indirect methods in \cite{Wal14, Ava11} alternatively finds the DPDFs using transformations of random variables (TRVs). In this way, the multiple random variables, to be averaged, can be reduced to a single random variable. Also, the indirect method can avoid the Fourier transform operation. In this way, the indirect method can produce a simpler DPSD solution than the solution obtained from the direct method. Yet, the conventional indirect method proposed in \cite{Ava11} assumes a statistical independence between angle-of-departure (AoD) and AoA, which is too strong assumption for the single bounced (SB) scattering considered in their model. Hence, their DPSD solution does not match to the DPSD, directly estimated from the channel gain function. Also, the analysis in \cite{Wal14} is not based on the channel gain function, linked to the geometrical model considered in their work. Accordingly, the proportionality validation between the delay-dependent DPDF and DPSD was infeasible. Besides, an arbitrary delay PDF needs to be assumed for the DPDF computation. Since an analytic channel gain function, corresponding to the DPDF, was not proposed in \cite{Wal14}, the research result is not directly applicable to fading simulator design. Finally, none of the above works analyzed the impact of the road layouts on the DPSD characteristics nor provided quantitative comparison between analytic, simulated, and measured DPSDs.} 

%
%
%
%
%

Bearing in mind the aforementioned limitations, we investigate the indirect method for the DPSD analysis of a generic 2D RSS model. At first, we formulate a stochastic channel gain function based on the geometrical model. To find an accurate and analytic DPSD of the channel gain function, we translate this problem into the problem of finding an analytic DPDF by using their equivalence as in \cite{Ava11} while further relaxing the independence assumption between AoD and AoA. 
%
In particular, we use \FBB{successive TRVs} to derive a closed-form joint AoD-AoA PDF and a joint Doppler-AoA PDF. We then marginalize the latter PDF over the AoA to obtain the DPDF, and hence DPSD. \FBB{In this way, the multiple integral-form solutions based on the conventional direct method \cite{Ava12} (triple integral) and indirect methods \cite{Ava11,Wal14} (double integral) can be reduced to a single integral. Furthermore, our indirect method presented herein does not require any additional assumptions on the AoD-AoA independence \cite{Ava11} and the delay PDF \cite{Wal14}, thereby more practical, accurate, and realistic for the analysis of the DPSD characteristics by RSSs, model validation (model parameter estimation) using measurement data, and efficient fading simulator design.}

It is noteworthy that all analytical results obtained in this paper are verified by simulation results. The closed-form joint AoD-AoA PDF and Doppler-AoA PDF are new results, and their properties are also investigated. Based on the new analytic DPSD solution, we investigate the impact of RSS layouts on the DPSD, Doppler spread, mean Doppler shift (MDS), and root-mean-square Doppler spread (RDS) for the first time in the literature. The DPSD and Doppler spread are compared to the modeled and measured DPSDs in \cite{Che13}. 
To validate our model, the new analytic DPSD is quantitatively compared to the measured DPSDs, collected for the IEEE 802.11p standard channels \cite{Aco07, Aco07_2}, via numerical optimizations. 

The rest of the paper is organized as follows. In Section II, the geometrical RSS model and channel gain function are introduced. In Section III, the new analytic DPSD is derived using the proposed method. The joint AoD-AoA PDF and joint Doppler-AoA PDF are derived in closed-forms, and the definitions of MDS and RDS are shown. In Section IV, all the analytic results are validated by simulations, and their properties are investigated. Also, the new DPSD is compared to the model and measured DPSDs in \cite{Che13}. In Section V, the new DPSD is compared to the measured DPSDs in \cite{Aco07, Aco07_2}. Finally, conclusions are drawn in Section VI. 
 
\section{Geometrical Roadside Scattering Model}

\begin{figure}[t]
    \centering
    \includegraphics [width=8.7cm] {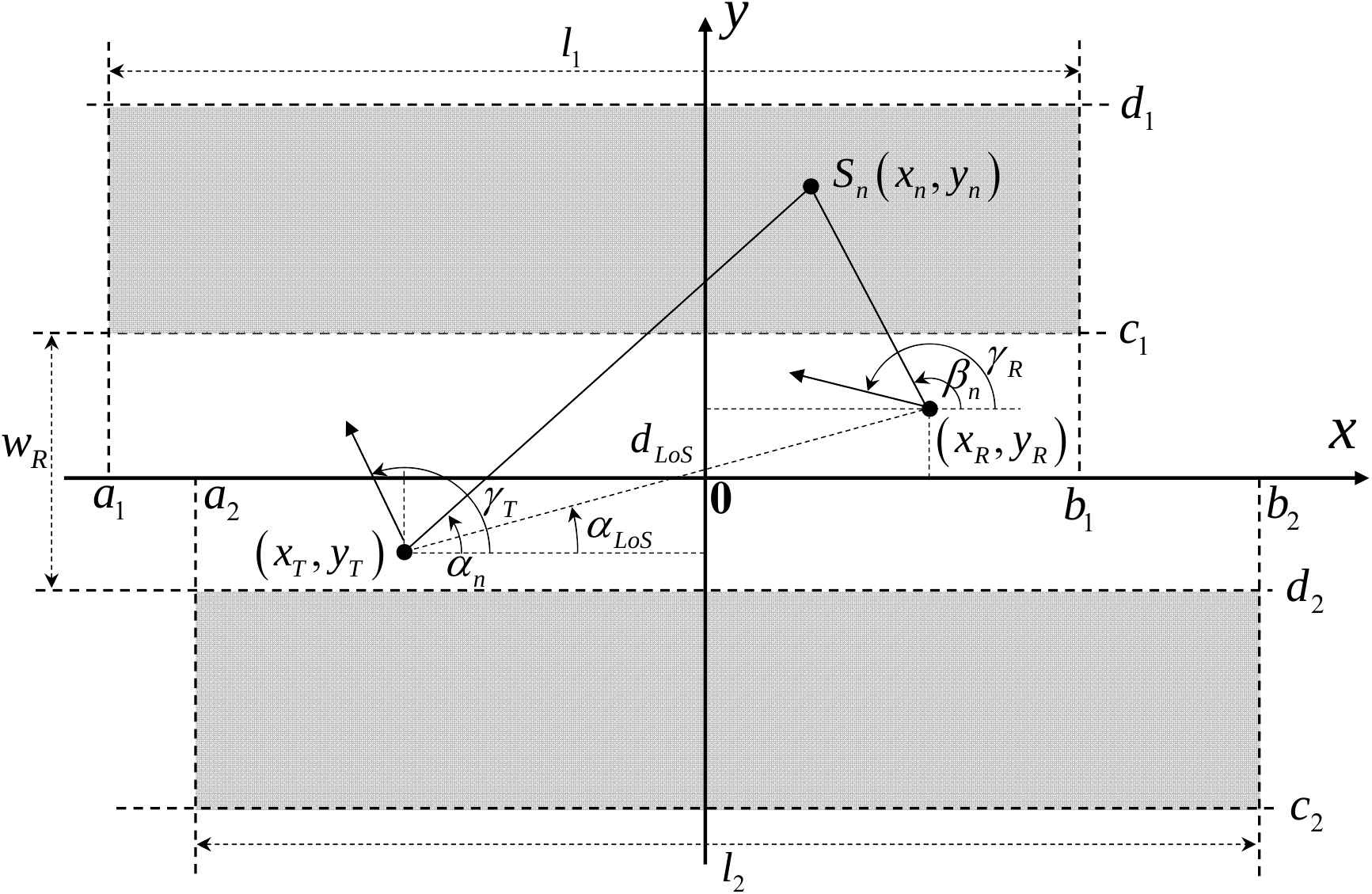}
    \caption{The geometric 2D RSS model for V2V communication channels.}
    \label{RSS_model}
\vspace{-0.5cm}
\end{figure}

The RSS model under consideration is presented in Fig. \ref{RSS_model}. It is assumed that the Tx and Rx vehicles are equipped with single isotropic antennas and move on the straight road on specific lanes, in the same direction (SD) or the opposite direction (OD). Also, the received signals are composed of a line-of-sight (LoS) component and SB components generated by the RSSs, so that the fading envelop follows  Rice distribution. The model geometry is similar to \cite{Kar09,Zho12, Ava12, Wal14, Ava11}, but is more general, allowing realistic asymmetric placement of two separate RSS regions. The model is represented in a 2D Cartesian coordinate system, where the location of a point is expressed by a pair of two real numbers, $(x,y)\in{\mathbb R^2}$. The $x$-axis is assumed as the middle lane of a road. The Tx and Rx are located at $(x_T, y_T)$ and $(x_R, y_R)$. They move with the velocities $v_T$ and $v_R$ in the directions determined by the angle of the motions $\gamma_T$ and $\gamma_R$, respectively. It is assumed that total $N$ number of stationary RSSs are uniformly distributed within the two shaded rectangular regions\footnote{In this paper, we assume that the average density (i.e., the number of the scatterers per a square meter) of RSSs is constant.}. 
The total RSS (shaded) region is defined by ${\cal B}={{\cal B}_1} \cup {{\cal B}_2}$, where  
\begin{IEEEeqnarray}{rCl}\label {eq:region_def}
{{\cal B}_i} = \left\{ {(x,y):{a_i} \le x \le {b_i} {~\rm and~} {c_i} \le y \le {d_i}} \right\}
\end{IEEEeqnarray}
denotes the upper RSS region for $i=1$ and the lower RSS region for $i=2$. The length of each region is $l_{i}=b_i-a_i$. The width of the road, identical to the minimum width of the unobstructed area, is defined as $w_R=c_1-d_2$. For $i\in\{1,2\}$, the model constraints are:
\begin{IEEEeqnarray}{rCl}\label {eq:MConst1}
\begin{array}{c}
{x_T} < {x_R},\\
{a_i}< {b_i},\\
{c_i} < {d_i},\\
\max \left\{ {{a_i}} \right\} < {x_T},\\
{x_R} < \min \left\{ {{b_i}} \right\},\\
\max \left\{ {{y_T},{y_R}} \right\} < {c_1},\\
{d_2} < \min \left\{ {{y_T},{y_R}} \right\}.
\end{array}
\end{IEEEeqnarray}
The constraints in (\ref{eq:MConst1}) are required to properly and realistically locate the two RSS regions w.r.t. the orientations of the Tx and Rx. They are also needed in the optimization problem design to estimate feasible model parameters from measurement data.

Based on the above definitions, $S_n$ denotes the $n$th RSS located at $\left(x_n, y_n\right)\in{\cal B}$, $n=1,2,...,N$, \FB{where $N=N_1 + N_2$ denotes the total number of scatterers. The model has $N_1$ scatterers in the upper and $N_2$ scatterers in the lower, where each scatterer is indexed as $S_{n_1}$ or $S_{n_2}$ with $n_i=1,2,...,N_i$ for $i\in\{1,2\}$. In this paper, we refer $S_{n_1}$ as an upper roadside scatterer (URS) and $S_{n_2}$ as a lower roadside scatterer (LRS), respectively.} 
By defining the sets of URSs and LRSs as ${\cal S}_{i}=\left\{S_{n_i}\right\}_{n_i=1}^{N_i}$ for $i\in\{1,2\}$, the total RSS set can be defined as ${\cal S} =  {\cal S}_1\cup {\cal S}_2$.
Under the uniform scattering density assumption, the coordinate of $S_{n} \in {{\cal B}}$, i.e., $(x_n,y_n)$, can be modeled by a pair of two independent uniform random variables, i.e., $(X_n, Y_n)$, identically characterized by a joint PDF for all $n$:
\begin{IEEEeqnarray}{rCl}\label {eq:XY_JPDF}
{f_{{X},{Y}}}\left( {x,y} \right) = \begin{array}{*{20}{c}}
{{A^{ - 1}} \cdot {{\bf{1}}_{{{\cal B}}}}}\left( {x,y} \right),
\end{array}
\end{IEEEeqnarray}
where $A{\rm{ = }}{A_1} + {A_2}$ with ${\rm{ }}{A_i} = \left( {{b_i} - {a_i}} \right)\left( {{d_i} - {c_i}} \right)$, and the indicator function in (\ref{eq:XY_JPDF}) is defined as:
\begin{IEEEeqnarray}{rcl}\label {eq:IF}
{{\bf{1}}_{{{\cal B}}}}\left( {x,y} \right) = \left\{ {\begin{array}{*{20}{c}}
{1,}&{{\rm{if }}\left( {x,y} \right) \in {{\cal B}}}\\
{0,}&{{\rm{otherwise}}}
\end{array}} \right..
\end{IEEEeqnarray}

In Fig. 1, $\alpha_n$ and $\beta_n$ denote the AoD and AoA, associated with $S_n$, respectively. From the model geometry and (\ref{eq:XY_JPDF})--(\ref{eq:IF}), it is clear that a pair of $n$th AoD and AoA, i.e., $\left(\alpha_n,\beta_n\right)$, solely depends on the random location of $S_n$, and hence, they are also random quantities depending on $\left(X_n,Y_n\right)$. We characterize $\left(\alpha_n,\beta_n\right)$ by a pair of two random variables, $\left({{\rm A_n}},{{\rm B_n}}\right)$, which are the piecewise functions of $X_n$ and $Y_n$ as below: 
%
\begin{IEEEeqnarray}{rcl}\label {eq:AoD}
{\rm A}_n = \left\{ {\begin{array}{*{20}{lll}}
{\arctan \left( {\frac{{{Y_n} - {y_T}}}{{{X_n} - {x_T}}}} \right),}&{{\rm{if ~ }}{X_n} > {x_T}}\\
{\arctan \left( {\frac{{{Y_n} - {y_T}}}{{{X_n} - {x_T}}}} \right) + \pi ,}&{{\rm{if~  }}{X_n} < {x_T}{\rm{, }}{Y_n} > {y_T}{\rm{ }}}\\
{\arctan \left( {\frac{{{Y_n} - {y_T}}}{{{X_n} - {x_T}}}} \right) - \pi ,}&{{\rm{if ~ }}{X_n} < {x_T}{\rm{, }}{Y_n} < {y_T}}
\end{array}} \right.,
\\\label {eq:AoA}
{\rm B}_n = \left\{ {\begin{array}{*{20}{lll}}
{\arctan \left( {\frac{{{Y_n} - {y_R}}}{{{X_n} - {x_R}}}} \right),}&{{\rm{if ~ }}{X_n} > {x_R}}\\
{\arctan \left( {\frac{{{Y_n} - {y_R}}}{{{X_n} - {x_R}}}} \right) + \pi ,}&{{\rm{if ~ }}{X_n} < {x_R}{\rm{, }}{Y_n} > {y_R}{\rm{ }}}\\
{\arctan \left( {\frac{{{Y_n} - {y_R}}}{{{X_n} - {x_R}}}} \right) - \pi ,}&{{\rm{if ~ }}{X_n} < {x_R}{\rm{, }}{Y_n} < {y_R}}
\end{array}} \right..
\end{IEEEeqnarray}

Assuming that the channel is wide-sense stationary (WSS), and based on the SOCE principle \FB{(see \cite{Che09_2} or pp. 60--61 of \cite{Patbook11})}, we model a normalized time-variant channel gain function of the geometrical RSS model as below:
%
%
\begin{IEEEeqnarray}{rCl}\label {eq:TotalCG}
h\left( t \right) = \sqrt{\frac{K}{{K + 1}}} h_{}^{{\rm{LoS}}}\left( t \right) + \sqrt{\frac{1}{ {K + 1}}}h_{}^{{\rm{RSS}}}\left( t \right).
\end{IEEEeqnarray}
\FB{Note that the model in (\ref{eq:TotalCG}) is a standard Rician fading model, where $K$ denotes the Rician $K$ factor for distributing the total power between the deterministic LoS component $h_{}^{{\rm{LoS}}}\left( t \right)$ and the diffuse component by RSSs $h_{}^{{\rm{RSS}}}\left( t \right)$. The LoS component is defined as below:}

%
\begin{IEEEeqnarray}{rCl}\label {eq:LoS_gain}
h_{}^{{\rm{LoS}}}\left( t \right) &=&{e^{j\left( {2\pi {f^{{\rm{LoS}}}}t - \frac{{2\pi }}{\lambda }{d_{{\rm{LoS}}}}} \right)}},
\end{IEEEeqnarray}
where $f^{\rm LoS}$ and $d_{\rm LoS}$ are the Doppler frequency of the LoS component, and the LoS distance, respectively, defined as:
%
\begin{IEEEeqnarray}{rcl}\label {eq:LoS_Doppler_freq}
{f^{{\rm{LoS}}}} &=& {f_{{T_{\max }}}}\cos \left( {{\alpha _{{\rm{LoS}}}} - {\gamma _T}} \right) + {f_{{R_{\max }}}}\cos \left(\pi+ {{\alpha _{{\rm{LoS}}}} - {\gamma _R}} \right),
\\\label{eq:LoS_distance}
{d_{{\rm{LoS}}}} &=& \sqrt {{{\left( {{x_R} - {x_T}} \right)}^2} + {{\left( {{y_R} - {y_T}} \right)}^2}},
\end{IEEEeqnarray}
where $f_{T_{\max}}=v_T/\lambda$ and $f_{R_{\max}}=v_R/\lambda$ denote the maximum Doppler frequencies due to the movements of the Tx and Rx, respectively. Here $\lambda=c_0/f_c$ is the wavelength, where $f_c$ and $c_0$ are the carrier frequency and the speed of light. ${\alpha _{{\rm{LoS}}}}$ in (\ref {eq:LoS_Doppler_freq}) denotes the AoD of the LoS component, defined as: 
\begin{IEEEeqnarray}{rcl}\label {eq:LoS_AoD}
{\alpha _{{\rm{LoS}}}} = \arctan \left( {{m_{{\rm{LoS}}}}} \right),
\end{IEEEeqnarray}
where ${m_{{\rm{LoS}}}}$ is the gradient of the LoS path:
\begin{IEEEeqnarray}{rcl}\label {eq:LoS_slope}
{m_{\rm LoS}} = \frac{{{y_R} - {y_T}}}{{{x_R} - {x_T}}}.
\end{IEEEeqnarray}
In (\ref{eq:TotalCG}), the diffuse component is modeled as: 
\begin{IEEEeqnarray}{rcl}\label{eq:RSS_gain} 
h_{}^{{\rm{RSS}}}\left( t \right) = \mathop {\lim }\limits_{N \to \infty } \sum\limits_{n = 1}^N {{\FB{g_n}}{e^{j\left( {{\Theta _n} + 2\pi {F_{D,n}}t} \right)}}},
\end{IEEEeqnarray}
where \FB{$g_n$} is the $n$th path gain, $\Theta_n$ is the random phase shift, modeled as independent, identically distributed (i.i.d.) uniform random variables, following $\cal{U}(-\pi, \pi)$ for all $n$. $F_{D,n}$ denotes the $n$th Doppler frequency by $S_n$, and is defined as:
\begin{IEEEeqnarray}{rCl}\label {eq:DF}
F_{D,n} = f_{T_{\max}}\cos({\rm A}_n -\gamma_T) +f_{R_{\max}}\cos({\rm B}_n -\gamma_R).
\end{IEEEeqnarray}

%
From (\ref{eq:AoD}), (\ref{eq:AoA}), and (\ref{eq:DF}), it is found that 1) the $n$th AoD and AoA are statistically dependent; and 2) the Doppler frequency $F_{D,n}$ is also a function of two random variables. Hence, the DPSD analysis of $h^{\rm RSS}\left(t\right)$ in (\ref{eq:RSS_gain}) must take into account the statistical dependency between ${\rm A}_n$ and ${\rm B}_n$. It is also noteworthy that the analytic solution of the DPSD in \cite{Ava11} is based on the independence between the AoD and AoA, not leading to exact DPSD shapes \FB{for SB scattering}.

\section{DPSD Analysis}
In this section, we describe the direct method for the derivation of the DPSD of $h\left(t\right)$ in (\ref{eq:TotalCG}), denoted as $S_{hh}\left(\nu\right)$. Then, an alternative indirect method is formulated,  the joint AoD-AoA PDF and joint Doppler-AoA PDF are derived in closed-forms, and we get a new $S_{hh}\left(\nu\right)$. In the end of the section, the definitions of MDS and RDS are represented for Rician channels. Note that both of the direct and the indirect methods assume the normalized equal path gain (EPG), i.e., ${\FB{g_n}}=1/{\sqrt N}$ in (\ref{eq:RSS_gain}) \cite{Hoe92}.

\subsection{Direct Method}
We start with the ACF definition of a WSS process $x(t)$:
\begin{IEEEeqnarray}{rcl}\label{eq:ACF_def}
R_{xx}\left( \tau  \right) = E\left[ {{x^*}\left( t \right)x\left( {t + \tau } \right)} \right],
\end{IEEEeqnarray}
where $(\cdot)^{*}$ denotes the complex conjugate. By substituting (\ref{eq:TotalCG}) into (\ref{eq:ACF_def}), we obtain
\begin{IEEEeqnarray}{rcl}\label{eq:ACF_result}
{R_{hh}}\left( \tau  \right) = \sqrt{\frac{K}{{K + 1}}}R_{hh}^{{\rm{LoS}}}\left( \tau  \right) + \sqrt{\frac{1}{{K + 1}}}R_{hh}^{{\rm{RSS}}}\left( \tau  \right),
\end{IEEEeqnarray}
where $R_{hh}^{{\rm{LoS}}}\left( \tau  \right)$ and $R_{hh}^{{\rm{RSS}}}\left( \tau  \right)$ refer to the ACFs of the normalized LoS and RSS components and are obtained as:
\begin{IEEEeqnarray}{rcl}
\label{eq:ACF_LoS}
R_{hh}^{{\rm{LoS}}}\left( \tau  \right) &=& {e^{j2\pi {f^{{\rm{LoS}}}}\tau }},\\\nonumber
R_{hh}^{{\rm{RSS}}}\left( \tau  \right) &=& \mathop {\lim }\limits_{N \to \infty } \frac{1}{N}\sum\limits_{n = 1}^N {E\left[ {{e^{j2\pi {F_{D,n}}\tau }}} \right]} {\rm{ }}\\
\label{eq:ACF_RSS1}
 &=& {\rm{ }}\int_{\nu  \in {\cal X}}^{} {{e^{j2\pi \nu \tau }}} {f_{F_D}}(\nu )d\nu\\
\label{eq:ACF_RSS2} 
& =& \sum\limits_{i = 1}^2 {A_i^{ - 1}\int_{y = {c_i}}^{{d_i}} {\int_{x = {a_i}}^{{b_i}} {{e^{j2\pi {F_D}\left( {x,y} \right)\tau }}} dxdy} } {\rm{ }}.
 \end{IEEEeqnarray}
Note that ${f_{F_{D}}}(\nu )$ in (\ref{eq:ACF_RSS1}) denotes the PDF of the Doppler frequencies $F_{D,n}$, which are i.i.d. $\forall n$, and $\cal X$ is the corresponding sample space. 
In the literature, the correct analytic expression of ${f_{F_{D}}}(\nu )$ have not been deduced. Instead, substituting (\ref{eq:DF}) into (\ref{eq:ACF_RSS1}) with the results in (\ref{eq:XY_JPDF})--(\ref{eq:AoA}), leads to (\ref{eq:ACF_RSS2}).

In order to obtain the DPSD of $h(t)$, the direct method takes a Fourier transform of (\ref{eq:ACF_result}) as below: 
\begin{IEEEeqnarray}{rcl}\nonumber
S_{hh}^{}\left( \nu  \right) &=& {{\cal F}_{\tau  \to \nu }}\left\{ {R_{hh}^{}\left( \tau  \right)} \right\}\\
\label{eq:PSD_total}
&=& \sqrt{\frac{K}{{K + 1}}}S_{hh}^{{\rm{LoS}}}\left( \nu  \right) + \sqrt{\frac{1}{{K + 1}}}S_{hh}^{{\rm{RSS}}}\left( \nu  \right),
\end{IEEEeqnarray}
where ${\cal F}\{\cdot\}$ denotes a Fourier transform operator. $S_{hh}^{{\rm{LoS}}}\left( \nu  \right)$ and $S_{hh}^{{\rm{RSS}}}\left( \nu  \right)$ denote the DPSDs of (\ref {eq:LoS_gain}) and (\ref{eq:RSS_gain}), respectively, and are obtained as:
%
\begin{IEEEeqnarray}{rcl}\label{eq:PSD_LoS}
S_{hh}^{{\rm{LoS}}}\left( \nu  \right) &=& \delta \left( {\nu  - {f^{{\rm{LoS}}}}} \right),\\
\label{eq:PSD_RSS}
S_{hh}^{{\rm{RSS}}}\left( \nu  \right) &=& \sum\limits_{i = 1}^2 {A_i^{ - 1}\int\limits_{ - \infty }^\infty  {\int\limits_{{c_i}}^{{d_i}} {\int\limits_{{a_i}}^{{b_i}} {{e^{j2\pi \left\{ {{F_D}\left( {x,y} \right) - \nu } \right\}\tau }}dxdyd\tau } } } } {\rm{ }}.
\end{IEEEeqnarray}
Similar with the eq. (33) of \cite{Ava12}, the direct method yields a triple integral-form for ${S_{hh}^{{\rm{RSS}}}\left( \nu  \right)}$ as in (\ref{eq:PSD_RSS}). 

\subsection{Indirect Method}
Our indirect method aims to derive a simpler form of $S_{hh}\left( \nu  \right)$
, by exploiting the following equality:
\begin{IEEEeqnarray}{rcl}\label{eq:PSD_PDF_EQ}
S_{hh}^{\rm RSS}\left(\nu\right) = f_{F_D}\left(\nu\right),
\end{IEEEeqnarray}
which holds if $\FB{g_n}=1/{\sqrt N}$ (see Appendix I of \cite{Hoe92}). To obtain $f_{F_D}\left(\nu\right)$, first, a joint AoD-AoA PDF ${f_{{\rm A},{\rm B}}}\left( {\alpha ,\beta } \right)$ is deduced, followed by a joint Doppler-AoA PDF ${f_{{F_D},{\rm B}}}(\nu ,\beta )$ via successive TRVs. By marginalizing ${f_{{F_D},{\rm B}}}(\nu ,\beta )$ over $\beta$, we obtain ${f_{{F_D}}}(\nu)$, which is equivalent to $S_{hh}^{\rm RSS}\left(\nu\right)$. Finally, substituting ${f_{{F_D}}}(\nu)$ into (\ref{eq:PSD_total}) leads to the new result of $S_{hh}\left(\nu\right)$.


%
%

\subsubsection{Derivation of the joint AoD-AoA PDF}
To derive $f_{\rm{A},{\rm B}}(\alpha, \beta)$, a TRVs from $(X_n,Y_n)$ into $({\rm A}_n, {\rm B}_n)$ is formed as below: 
\begin{IEEEeqnarray}{rcl}\label {eq:JAoDAoA}
{f_{{\rm A},{\rm B}}}\left( {\alpha ,\beta } \right) = {A^{ - 1}}{{\bf{1}}_{\cal B}}\left( {x,y} \right)\left| {J(\alpha ,\beta )} \right|.
\end{IEEEeqnarray}
In (\ref{eq:JAoDAoA}), $n$ is omitted due to the i.i.d. property. From (\ref{eq:AoD}) and (\ref{eq:AoA}), $x$ and $y$ can be expressed as functions of $\alpha$ and $\beta$ as: 
\begin{IEEEeqnarray}{cll}\label {eq:IMAP1}
x &=& \frac{{{x_T}\tan {\alpha} - {x_R}\tan {\beta} + {y_R} - {y_T}}}{{\tan {\alpha} - \tan {\beta}}},\\ 
\label {eq:IMAP2}
y &=& \frac{{\left( {{x_T} - {x_R}} \right)\tan {\alpha}\tan {\beta} - {y_T}\tan {\beta} + {y_R}\tan {\alpha}}}{{\tan {\alpha} - \tan {\beta}}}.
\end{IEEEeqnarray}
Using (\ref{eq:IMAP1}) and (\ref{eq:IMAP2}), the Jacobian ${J({\alpha },{\beta })}$ in (\ref{eq:JAoDAoA}) is given by
%
\begin{IEEEeqnarray}{rcl}\label {eq:Jacobian1}
\begin{array}{l}
{J({\alpha},{\beta})}  = \left| {\begin{array}{*{20}{c}}
{\frac{{\partial {x}}}{{\partial {\alpha}}}}&{\frac{{\partial {x}}}{{\partial {\beta}}}}\\
{\frac{{\partial {y}}}{{\partial {\alpha}}}}&{\frac{{\partial {y}}}{{\partial {\beta}}}}
\end{array}} \right|\\
 {~~~~~~~~~}=  { - {{({x_T} - {x_R})}^2}\csc^3 {{\left( {{\alpha} - {\beta}} \right)}}}
{ \left( {\sin {\alpha} - {m_{\rm LoS}}\cos {\alpha}} \right)\left( {\sin {\beta} - {m_{\rm LoS}}\cos {\beta}} \right)}.
\end{array}
\end{IEEEeqnarray}
%
%
By substituting (\ref{eq:Jacobian1}) into (\ref{eq:JAoDAoA}), a closed-form expression of ${f_{{\rm A},{\rm B}}}\left( {\alpha ,\beta } \right)$ is obtained as:
\begin{IEEEeqnarray}{rcl}\label {eq:JAoDAoA2}
\begin{array}{l}
{f_{{\rm A},{\rm B}}}\left( {\alpha ,\beta } \right) = {A^{ - 1}}{{\bf{1}}_{\cal A}\left(\alpha,\beta \right)} \left| { {{({x_T} - {x_R})}^2}\csc^3 {{\left( {{\alpha} - {\beta}} \right)}}} \right.\\
\left. { {~~~~~~~~~~~~}\times\left( {\sin {\alpha} - {m_{\rm LoS}}\cos {\alpha}} \right)\left( {\sin {\beta} - {m_{\rm LoS}}\cos {\beta}} \right)} \right|,
\end{array}
\end{IEEEeqnarray}
\begin{figure}[t]
    \centering
    \includegraphics [width=8.7cm] {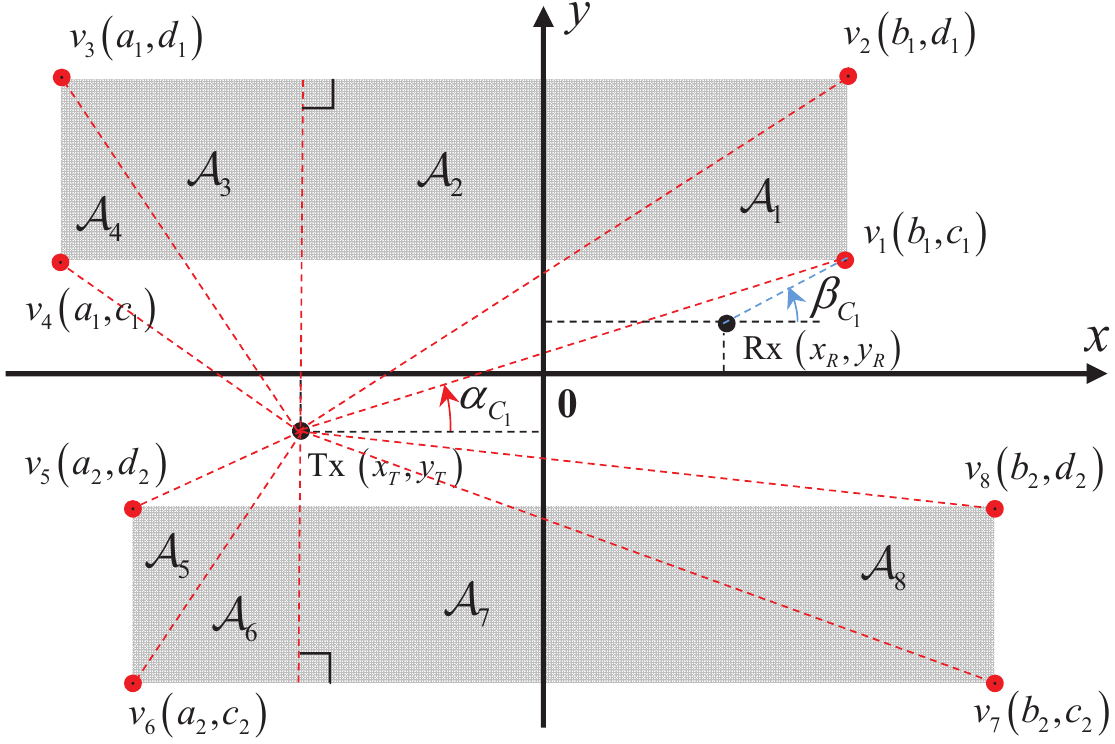}
    \caption{A geometrical representation of the subsample spaces ${\cal A}_k$, CAoDs $\alpha_{C_r}$, and CAoAs $\beta_{C_r}$. $v_r$ denotes the $r$th vertex of the total RSS region, $\cal B$.}
    \label{Subspace}
\vspace{-0.5cm}
\end{figure}
where ${\cal A} = \bigcup\nolimits_{k = 1}^{K=8} {{{\cal A}_k}}$ is the joint sample space of $\left({\rm A}_n,{\rm B}_n\right), \forall n$. Here ${{\cal A}_k}$ is a subsample space defined in (\ref{eq:Samplespace1}), shown in the next page. In (\ref{eq:Samplespace1}), ${{\rm{atan2}}\left(y,x \right)}$ is the four-quadrant inverse tangent function, returning angles within $(-\pi, \pi]$. $m_q$ for $q\in\{1,2,...,16\}$ are constants defined by the model geometry in Fig. \ref{RSS_model}. The explicit expressions are given in (\ref{eq:m_const}), presented in the next page. 
%
Parameter $\alpha_{C_r}$ denotes the critical AoD (CAoD) at the $r$th vertex of the RSS region $\cal B$, and is defined in (\ref{eq:CAoD}) for $r\in\{1,2,...8\}$, shown in the next page. The associated critical AoAs (CAoAs), $\beta_{C_r}$ for $r\in\{1,2,...8\}$ are defined similarly as in (\ref{eq:CAoD}), but $x_T$ and $y_T$ have to be replaced with $x_R$ and $y_R$ for all $r$. In Fig. \ref{Subspace}, ${{\cal A}_k}$, $\alpha_{C_r}$, and $\beta_{C_r}$ are visualized. In Fig. \ref{Subspace}, red dots refer to the vertices of the two rectangular RSS regions. The $k$th subsample space ${{\cal A}_k}$ is denoted in the corresponding shaded region, defined by (\ref{eq:Samplespace1}).
%
Note that the lower and upper bounds of the AoA $\beta$ in (\ref{eq:Samplespace1}) are correct only if $\alpha _{C_8}<{\alpha_{\rm LoS}} < \alpha _{C_1}$. 
Otherwise, if ${\alpha _{{C_1}}} < {\alpha_{\rm LoS}}<{\alpha _{{C_2}}}$, the upper and lower bounds of $\beta$ in ${{\cal A}_k}$ for $k\in\{1,5\}$ should be switched. If ${\alpha _{{C_2}}} < {\alpha_{\rm LoS}}<\pi/2$, the upper and lower bounds of $\beta$ in ${{\cal A}_k}$ for $k\in\{1,2,5,6\}$ should be switched. 
%

%
%
%
\begin{sidewaysfigure}
\normalsize

\begin{IEEEeqnarray}{l}
\arraycolsep=5pt\def\arraystretch{0.1}
\label{eq:Samplespace1}
\begin{array}{l}
{{\cal A}_1} \in \left\{ {\left( {\alpha ,\beta } \right):{\alpha _{{C_1}}} \le \alpha  < {\alpha _{{C_2}}}{\rm{, ~}}\arctan \left( {{m_1}\tan \alpha  + {m_2}} \right) \le \beta  \le {\rm{atan2}}\left( {\tan \alpha ,{m_3}\tan \alpha  + {m_4}} \right)} \right\},
\\
{{\cal A}_2} \in \left\{ {\left( {\alpha ,\beta } \right):{\alpha _{{C_2}}} \le \alpha  < \frac{\pi }{2}{\rm{,~~~ atan2}}\left( {\tan \alpha ,{m_5}\tan \alpha  + {m_6}} \right) \le \beta  \le {\rm{atan2}}\left( {\tan \alpha ,{m_3}\tan \alpha  + {m_4}} \right)} \right\},
\\
{{\cal A}_3} \in \left\{ {\left( {\alpha ,\beta } \right):\frac{\pi }{2} \le \alpha  \le {\alpha _{{C_3}}}{\rm{,~~~ atan2}}\left( {\tan \alpha ,{m_5}\tan \alpha  + {m_6}} \right) + \pi  \le \beta  \le {\rm{atan2}}\left( {\tan \alpha ,{m_3}\tan \alpha  + {m_4}} \right) + \pi } \right\},
\\
{{\cal A}_4} \in \left\{ {\left( {\alpha ,\beta } \right):{\alpha _{{C_3}}} \le {\alpha} \le {\alpha _{{C_4}}}{\rm{, ~}}\arctan \left( {{m_7}\tan \alpha  + {m_8}} \right) + \pi  \le \beta  \le {\rm{atan2}}\left( {\tan \alpha ,{m_3}\tan \alpha  + {m_4}} \right) + \pi } \right\},
\\
{{\cal A}_5} \in \left\{ {\left( {\alpha ,\beta } \right):{\alpha _{{C_5}}} \le \alpha  < {\alpha _{{C_6}}}{\rm{,~~ atan2}}\left( {\tan \alpha ,{m_9}\tan \alpha  + {m_{10}}} \right) - \pi  \le \beta  \le \arctan \left( {{m_{11}}\tan \alpha  + {m_{12}}} \right) - \pi } \right\},
\\
{{\cal A}_6} \in \left\{ {\left( {\alpha ,\beta } \right):{\alpha _{{C_6}}} \le \alpha  <  - \frac{\pi }{2}{\rm{, ~~atan2}}\left( {\tan \alpha ,{m_9}\tan \alpha  + {m_{10}}} \right) - \pi  \le \beta  \le {\rm{atan2}}\left( {\tan \alpha ,{m_{13}}\tan \alpha  + {m_{14}}} \right) - \pi } \right\},
\\
{{\cal A}_7} \in \left\{ {\left( {\alpha ,\beta } \right): - \frac{\pi }{2} \le \alpha  < {\alpha _{{C_7}}}{\rm{,~~ atan2}}\left( {\tan \alpha ,{m_9}\tan \alpha  + {m_{10}}} \right) \le \beta  \le {\rm{atan2}}\left( {\tan \alpha ,{m_{13}}\tan \alpha  + {m_{14}}} \right)} \right\},
\\
{{\cal A}_8} \in \left\{ {\left( {\alpha ,\beta } \right):{\alpha _{{C_7}}} \le \alpha  < {\alpha _{C8}}{\rm{, ~~atan2}}\left( {\tan \alpha ,{m_9}\tan \alpha  + {m_{10}}} \right) \le \beta  \le \arctan \left( {{m_{15}}\tan \alpha  + {m_{16}}} \right)} \right\}, {\rm where}
\end{array}
\\
%
\arraycolsep=5pt\def\arraystretch{1}
\begin{array}{*{20}{l}}\label{eq:m_const}
{{m_1} = \frac{{{b_1} - {x_T}}}{{{b_1} - {x_R}}},}&
{{m_2} = \frac{{{y_T} - {y_R}}}{{{b_1} - {x_R}}},}&
{{m_3} = \frac{{{x_T} - {x_R}}}{{{c_1} - {y_R}}},}&
{{m_4} = \frac{{{c_1} - {y_T}}}{{{c_1} - {y_R}}},}&
{{m_5} = \frac{{{x_T} - {x_R}}}{{{d_1} - {y_R}}},}&
{{m_6} = \frac{{{d_1} - {y_T}}}{{{d_1} - {y_R}}},}&
{{m_7} = \frac{{{a_1} - {x_T}}}{{{a_1} - {x_R}}},}\\
{{m_8} = \frac{{{y_T} - {y_R}}}{{{a_1} - {x_R}}},}&
{{m_9} = \frac{{{x_T} - {x_R}}}{{{d_2} - {y_R}}},}&
{{m_{10}} = \frac{{{d_2} - {y_T}}}{{{d_2} - {y_R}}},}&
{{m_{11}} = \frac{{{a_2} - {x_T}}}{{{a_2} - {x_R}}},}&
{{m_{12}} = \frac{{{y_T} - {y_R}}}{{{a_2} - {x_R}}},}&
{{m_{13}} = \frac{{{x_T} - {x_R}}}{{{c_2} - {y_R}}},}&
{{m_{14}} = \frac{{{c_2} - {y_T}}}{{{c_2} - {y_R}}},}\\
{{m_{15}} = \frac{{{b_2} - {x_T}}}{{{b_2} - {x_R}}},}&
{{m_{16}} = \frac{{{y_T} - {y_R}}}{{{b_2} - {x_R}}},}&
{{m_{17}} = \frac{{{{d_1} - {y_T}}}}{{{{b_1} - {x_T}}}},}&
{{m_{18}} = \frac{{{d_1} - {y_T}}}{{{a_1} - {x_T}}},}&
{{m_{19}} = \frac{{{c_2} - {y_T}}}{{{a_2} - {x_T}}},}&
{{m_{20}} = \frac{{{c_2} - {y_T}}}{{{b_2} - {x_T}}}.}\\
\end{array}
\\
\arraycolsep=5pt\def\arraystretch{1}
\begin{array}{*{20}{l}}\label{eq:CAoD}
{{\alpha _{{C_1}}} = \arctan \left( {\frac{{{c_1} - {y_T}}}{{{b_1} - {x_T}}}} \right),}&{{\alpha _{{C_2}}} = \arctan \left( {\frac{{{d_1} - {y_T}}}{{{b_1} - {x_T}}}} \right),}&{{\alpha _{{C_3}}} = \arctan \left( {\frac{{{d_1} - {y_T}}}{{{a_1} - {x_T}}}} \right) + \pi ,}&{{\alpha _{{C_4}}} = \arctan \left( {\frac{{{c_1} - {y_T}}}{{{a_1} - {x_T}}}} \right) + \pi ,}\\
{{\alpha _{{C_5}}} = \arctan \left( {\frac{{{d_2} - {y_T}}}{{{a_2} - {x_T}}}} \right) - \pi ,}&{{\rm{ }}{\alpha _{{C_6}}} = \arctan \left( {\frac{{{c_2} - {y_T}}}{{{a_2} - {x_T}}}} \right) - \pi ,}&{{\alpha _{{C_7}}} = \arctan \left( {\frac{{{c_2} - {y_T}}}{{{b_2} - {x_T}}}} \right),}&{{\alpha _{{C_8}}} = \arctan \left( {\frac{{{d_2} - {y_T}}}{{{b_2} - {x_T}}}} \right).}
\end{array}
\\\nonumber
\hrulefill\\
\arraycolsep=5pt\def\arraystretch{0.01}
\begin{array}{l}\label{eq:Samplespace2}
\small
{{\cal D}_1} \in \left\{ {\begin{array}{*{20}{c}}
{\left( {\nu ,\beta } \right):{f_R}\left( \beta  \right) + {f_T}\left( {\frac{{{m_2} - \tan \beta }}{{{m_1}}}} \right) \le \nu  \le {f_R}\left( \beta  \right) + {f_T}\left( {\frac{{{m_4}\tan \beta }}{{{m_3}\tan \beta  - 1}}} \right){\rm{,}}}&{\arctan {\beta _{{C_1}}} \le \beta  \le {\rm{atan2}}\left( {{m_{17}},{m_3}{m_{17}} + {m_4}} \right) + \pi {{\bf{1}}_{[\frac{\pi }{2},\pi ]}}\left( \beta  \right)}
\end{array}} \right\}\\
\small
{{\cal D}_2} \in \left\{ {\begin{array}{*{20}{c}}
{\left( {\nu ,\beta } \right):{f_R}\left( \beta  \right) + {f_T}\left( {\frac{{{m_6}\tan \beta }}{{{m_5}\tan \beta  - 1}}} \right) \le \nu  \le {f_R}\left( \beta  \right) + {f_T}\left( {\frac{{{m_4}\tan \beta }}{{{m_3}\tan \beta  - 1}}} \right),}&{\arctan {\beta _{{C_2}}} \le \beta  \le {\rm{atan2}}\left( {1,{m_3}} \right) + \pi }
\end{array}} \right\}\\
\small
{{\cal D}_3} \in \left\{ {\begin{array}{*{20}{c}}
{\left( {\nu ,\beta } \right):{f_R}\left( \beta  \right) - {f_T}\left( {\frac{{{m_6}\tan \beta }}{{{m_5}\tan \beta  - 1}}} \right) \le \nu  \le {f_R}\left( \beta  \right) - {f_T}\left( {\frac{{{m_4}\tan \beta }}{{{m_3}\tan \beta  - 1}}} \right){\rm{, }}}&{{\rm{atan2}}\left( {1,{m_5}} \right) + \pi  \le \beta  \le {\rm{atan2}}\left( {{m_{18}},{m_3}{m_{18}} + {m_4}} \right) + \pi }
\end{array}} \right\}\\
\small
{{\cal D}_4} \in \left\{ {\begin{array}{*{20}{c}}
{\left( {\nu ,\beta } \right):{f_R}\left( \beta  \right) - {f_T}\left( {\frac{{{m_8} - \tan \beta }}{{{m_7}}}} \right) \le \nu  \le {f_R}\left( \beta  \right) - {f_T}\left( {\frac{{{m_4}\tan \beta }}{{{m_3}\tan \beta  - 1}}} \right){\rm{,}}}&{~\arctan \beta_{C_3} + \pi  \le \beta  \le \arctan \beta_{C_4}+ \pi }
\end{array}{\rm{ }}} \right\}\\
\small
{{\cal D}_5} \in \left\{ {\begin{array}{*{20}{c}}
{\left( {\nu ,\beta } \right):{f_R}\left( \beta  \right) - {f_T}\left( {\frac{{{m_{10}}\tan \beta }}{{{m_9}\tan \beta  - 1}}} \right) \le \nu  \le {f_R}\left( \beta  \right) - {f_T}\left( {\frac{{{m_{12}} - \tan \beta }}{{{m_{11}}}}} \right){\rm{,}}}&{{\rm{ }}~\arctan \beta_{C_5} - \pi  \le \beta  \le \arctan \beta_{C_6} - \pi }
\end{array}} \right\}\\
\small
{{\cal D}_6} \in \left\{ {\begin{array}{*{20}{c}}
{\left( {\nu ,\beta } \right):{f_R}\left( \beta  \right) - {f_T}\left( {\frac{{{m_{10}}\tan \beta }}{{{m_9}\tan \beta  - 1}}} \right) \le \nu  \le {f_R}\left( \beta  \right) - {f_T}\left( {\frac{{{m_{14}}\tan \beta }}{{{m_{13}}\tan \beta  - 1}}} \right){\rm{,}}}&{{\rm{ atan2}}\left( {{m_{19}},{m_9}{m_{19}} + {m_{10}}} \right) - \pi  \le \beta  \le {\rm{atan2}}\left( {1,{m_{13}}} \right) - \pi }
\end{array}} \right\}\\
\small
{{\cal D}_7} \in \left\{ {\begin{array}{*{20}{c}}
{\left( {\nu ,\beta } \right):{f_R}\left( \beta  \right) + {f_T}\left( {\frac{{{m_{10}}\tan \beta }}{{{m_9}\tan \beta  - 1}}} \right) \le \nu  \le {f_R}\left( \beta  \right) + {f_T}\left( {\frac{{{m_{14}}\tan \beta }}{{{m_{13}}\tan \beta  - 1}}} \right){\rm{,}}}&{{\rm{atan2}}\left( {1,{m_9}} \right) - \pi  \le \beta  \le \arctan {\beta _{{C_7}}}}
\end{array}{\rm{ }}} \right\}\\
\small
{{\cal D}_8} \in \left\{ {\begin{array}{*{20}{c}}
{\left( {\nu ,\beta } \right):{f_R}\left( \beta  \right) + {f_T}\left( {\frac{{{m_{10}}\tan \beta }}{{{m_9}\tan \beta  - 1}}} \right) \le \nu  \le {f_R}\left( \beta  \right) + {f_T}\left( {\frac{{{m_{16}} - \tan \beta }}{{{m_{15}}}}} \right){\rm{,}}}&{{\rm{atan2}}\left( {{m_{20}},{m_9}{m_{20}} + {m_{10}}} \right) - \pi {{\bf{1}}_{( - \pi , - \frac{\pi }{2}]}}\left( \beta  \right) \le \beta  \le \arctan {\beta _{{C_8}}}}
\end{array}{\rm{ }}} \right\}
\end{array}
\end{IEEEeqnarray}
\vspace*{-2pt}
\end{sidewaysfigure}
\subsubsection{Derivation of the joint Doppler-AoA PDF}
Using the joint AoD-AoA PDF in (\ref{eq:JAoDAoA2}) and the forward mapping in (\ref{eq:DF}), a TRVs from $(\rm{A_n}, \rm{B_n})$ to $(F_{D,n}, \rm{B}_n)$ is performed as below: 
\begin{IEEEeqnarray}{rcl}\label {eq:JDoppler-AoA}
{f_{{F_D},{\rm B}}}(\nu ,\beta ) = {f_{{\rm A},{\rm B}}}\left( {{h_1}\left( {\nu ,\beta } \right),\beta } \right) \cdot \left| {J(\nu ,\beta )} \right|,
\end{IEEEeqnarray}
where the inverse mapping is obtained from (\ref{eq:DF}) as: 
\begin{IEEEeqnarray}{rcl}\nonumber
&\alpha &= {h_1}\left( {\nu ,\beta } \right) \\
\label {eq:InverseDoppler}
&&= {\left( { - 1} \right)^{i - 1}} \cdot \arccos \left\{ {z\left( \nu, \beta  \right)} \right\} + {\gamma _T},\\
&z\left( \nu,  \beta \right)&= \left\{ {\nu  - {f_{{R_{\max }}}}\cos \left( {\beta  - {\gamma _R}} \right)} \right\}f_{{T_{\max }}}^{ - 1}.
\end{IEEEeqnarray}
%
%
%
%
%
In (\ref{eq:InverseDoppler}), $i=1$ if $\beta_{C_1}\le \beta \le \beta_{C_4}$ (or equivalently, $S_n\in {{\cal S}_1}$); otherwise, $i=2$. 
The Jacobian is given as: 
\begin{IEEEeqnarray}{rcl}\label {eq:Jacobian_AoD}
%
J\left( {\nu ,\beta } \right) = \frac{\partial }{{\partial {\nu}}}{h_1}\left( {\nu ,\beta } \right) = \frac{{{{\left( { - 1} \right)}^{i-1}}}}{{{f_{{T_{\max }}}}\sqrt {1 - {z^2}\left( \nu, \beta  \right)} }}.
\end{IEEEeqnarray}
By substituting (\ref{eq:InverseDoppler}--\ref{eq:Jacobian_AoD}) into (\ref{eq:JDoppler-AoA}), we obtain: 
%
\begin{IEEEeqnarray}{rcl}\label {eq:joint_Doppler_AoA_PDF}\nonumber
{f_{{F_D},{\rm B}}}\left( {\nu ,\beta } \right) &=& \frac{{{{({x_T} - {x_R})}^2}}}{{A{f_{{T_{\max }}}}}}{{\bf{1}}_{\cal D}}\left( {\nu ,\beta } \right)\\\nonumber
&\times& \left| {\frac{{{{\csc }^3}\left( {{h_1}\left( {\nu ,\beta } \right) - \beta } \right)}}{{\sqrt {1 - {z^2}\left( {\nu ,\beta } \right)} }}\left[ {\sin \left\{ {{h_1}\left( {\nu ,\beta } \right)} \right\} - {m_{{\rm{LoS}}}}\cos \left\{ {{h_1}\left( {\nu ,\beta } \right)} \right\}} \right]\left( {\sin \beta  - {m_{{\rm{LoS}}}}\cos \beta } \right)} \right|\\
\end{IEEEeqnarray}
where ${\cal D} = \bigcup\nolimits_{k = 1}^{K = 8} {{{\cal D}_k}} $ is the sample space of $\left(F_{D,n}, {\rm B}_n\right), \forall n$. ${{\cal D}_k}$ is a subsample space defined in (\ref{eq:Samplespace2}), shown in the page $12$, and corresponds to the subsample space ${\cal A}_k$ in Fig. \ref{Subspace}. 
In (\ref{eq:Samplespace2}), ${f_T}\left( x \right)$ and ${\rm{ }}{f_R}\left( \beta  \right)$ denote 
the Tx and Rx Doppler frequencies w.r.t. AoA: 
\begin{IEEEeqnarray}{rcl}\label {eq:DF_AoA1}
{f_T}\left( x \right)  &=& {f_{{T_{\max }}}}\cos \left( {{\gamma _T} + \arctan \left( x \right)} \right),
\\\label {eq:DF_AoA2}
{f_R}\left( \beta  \right) &=& {f_{{R_{\max }}}}\cos \left( {\beta  - {\gamma _R}} \right).
\end{IEEEeqnarray}
Note that $m_q$ for $q\in\{1,2,...,20\}$ are constants associated with the model geometry and defined in (\ref{eq:m_const}), presented in the page $12$.
%

%

To explain how we obtained ${{\cal D}_k}$ in (\ref{eq:Samplespace2}), let us denote $f _{\min}^{{\cal A}_k}(\beta)$ and $f _{\max}^{{\cal A}_k}(\beta)$ as the lower and upper bounds of the Doppler frequency $\nu$ in the $k$th subspace ${\cal D}_k$. Further, $\beta _{\min}^{{\cal A}_k}$ and $\beta _{\max}^{{\cal A}_k}$ are similarly defined for the AoA $\beta$ in ${\cal D}_k$. Then, we can rewrite ${{\cal D}_k}$ in (\ref{eq:Samplespace2}) for $k\in\{1,2,...,8\}$ as below:
%
\begin{IEEEeqnarray}{rcl}\label {eq:Dk}
{{\cal D}_k} \in \left\{ {\left( {\nu ,\beta } \right):f _{\min}^{{\cal A}_k}\left( \beta  \right) \le \nu  \le f _{\max}^{{\cal A}_k}\left( \beta  \right){\rm{, }}\beta _{\min}^{{\cal A}_k} \le \beta  \le \beta _{\max}^{{\cal A}_k}} \right\}.
\end{IEEEeqnarray}
For ${\cal D}_k$, $f _{\min}^{{\cal A}_k}(\beta)$ and $f _{\max}^{{\cal A}_k}(\beta)$ 
can be found by substituting (\ref{eq:InverseDoppler}) into the lower and upper bounds of $\beta$ in ${\cal A}_k$, see (\ref{eq:Samplespace1}), and solving the resulting expressions for $\nu$. 
Since $f _{\min}^{{\cal A}_k}(\beta)$ and $f _{\max}^{{\cal A}_k}(\beta)$ are valid for the range of $\alpha$ in ${\cal A}_k$, $\beta _{\min}^{{\cal A}_k}$ and $\beta _{\max}^{{\cal A}_k}$ are the minimum and maximum AoA values in ${\cal A}_k$, respectively. The expressions can be obtained by investigating the geometry in Fig. \ref{Subspace}. 
%
%
Similar to the ${\cal A}_k$ case, it is important to emphasize that  the bounds of $\nu$ in (\ref{eq:Samplespace2}) are correct only if $\alpha _{C_8}<{\alpha_{\rm LoS}} < \alpha _{C_1}$. Otherwise, if ${\alpha _{{C_1}}} < {\alpha_{\rm LoS}}<{\alpha _{{C_2}}}$, $f _{\min}^{{\cal A}_k}(\beta)$ and $f _{\max}^{{\cal A}_k}(\beta)$ should be switched for $k\in\{1,5\}$. If ${\alpha _{{C_2}}} < {\alpha_{\rm LoS}}<\pi/2$, $f _{\min}^{{\cal A}_k}(\beta)$ and $f _{\max}^{{\cal A}_k}(\beta)$ should be switched for $k\in\{1,2,5,6\}$.

It is important to emphasize that the intervals of $\beta$, i.e., $\beta _{\min }^{{{\cal A}_k}} \le \beta  \le \beta _{\max }^{{{\cal A}_k}}$ for $k\in\{1,2,...,8\}$, can be overlapped depending on the choice of the model parameters, i.e., ${x_T}$,${y_T}$,${x_R}$,${y_R}$, $a_i$, $b_i$, $c_i$, $d_i$, $\forall i$. This eventually leads to $\bigcap\nolimits_{k = 1}^8 {{{\cal D}_k}}  \ne \phi$. Hence, finding the mutually disjoint subsample space, i.e., ${{\cal E}_u}$ satisfying ${\cal D} = \bigcup\nolimits_{u = 1}^U {{{\cal E}_u}}$ and $\bigcap\nolimits_{u = 1}^U {{{\cal E}_u}}  = \phi$, is important for the correct analysis of the Doppler-AoA PDF and marginal DPDF. Note that ${\cal E}_u$ can be easily found by sorting $\{\beta _{\min }^{{{\cal A}_k}}\}_{k=1}^{8}$ and $\{\beta _{\max }^{{{\cal A}_k}}\}_{k=1}^{8}$, respectively, formulating disjoint intervals of $\beta$, and finally assigning correct Doppler frequency bounds, $f _{\min}^{{\cal A}_k}(\beta)$ and $f _{\max}^{{\cal A}_k}(\beta)$, on the disjoint intervals. 
\subsubsection{DPDF}
The DPDF $f_{F_D}\left(\nu\right)$ can be obtained by marginalizing ${f_{{F_D},{\rm B}}}(\nu ,\beta )$ in (\ref{eq:joint_Doppler_AoA_PDF}) over $\beta$ as below: 
\begin{IEEEeqnarray}{rcl}\label{eq:Doppler_PDF}
f_{{F_D}}^{}\left( \nu  \right) = \int_\beta ^{} {{f_{{F_D},{\rm B}}}\left( {\nu ,\beta } \right)d\beta }.
\end{IEEEeqnarray}
Note that ${f_{{F_D},{\rm B}}}(\nu ,\beta )$ given in (\ref{eq:joint_Doppler_AoA_PDF}) includes the indicator function ${{\bf{1}}_{\cal D}}\left( {\nu,\beta } \right)$, which specifies the bounds of $\beta$ for a given Doppler frequency $\nu$. The numerical evaluation of (\ref{eq:Doppler_PDF}) requires the upper and lower bounds of $\beta$ for a given $\nu$, and those can be numerically computed from the inverses of $f _{\min}^{{\cal A}_k}(\beta)$ and $f _{\max}^{{\cal A}_k}(\beta)$ in (\ref{eq:Samplespace2}) by applying spline interpolations. 


\subsubsection{New analytic DPSD}
Based on (\ref{eq:PSD_PDF_EQ}) and by substituting (\ref{eq:PSD_LoS}) and (\ref{eq:Doppler_PDF}) into (\ref{eq:PSD_total}), we obtain the {\it ``new analytic DPSD''} as below: 
\begin{IEEEeqnarray}{rcl}\label{Proposed_PSD}
{S_{hh}}\left( \nu  \right) =\sqrt{\frac{K}{{K + 1}}} \delta \left( {\nu  - {f^{{\rm{LoS}}}}} \right) + \sqrt{\frac{1}{{K + 1}}}f_{{F_D}}^{}\left( \nu  \right),
\end{IEEEeqnarray}
%
%
%
which is a weighted sum of a Dirac delta function and the DPDF. These two components characterize a deterministic Doppler shift of the LoS path and random Doppler frequency shifts by RSSs, respectively. Since the area of the PDF is equal to $1$, $S_{hh}\left(\nu\right)$ is a normalized DPSD. 

It is noteworthy that the range of the Doppler frequency $\nu$ of $S_{hh}^{}\left( \nu  \right)$ can be smaller than those predicted by the RS-GBSCMs \cite{Akk86, Pat05, Zaj08, Che09_2, Zaj14, Zaj15, XChe13, Zaj09, Yua14} and the 1D RSS model of \cite{Che13}. For instance, in a SD scenario ($\gamma_T=\gamma_R=0$), the DPSDs of the mentioned models span over $|\nu| \le f_{D_{\max}}$, where $f_{D_{\max}}={f_{{T_{\max }}}} + {f_{{R_{\max }}}}$ denotes the maximum possible Doppler frequency. In real road environments, however, the RSS lengths can be bounded due geographical limits and path-loss. Hence, the AoD and AoA ranges can be reduced to smaller than $2\pi$. The Doppler range of our RSS model is bounded as:
\begin{IEEEeqnarray}{rcl}\label{eq:Doppler_range1}
-f_{D_{\max}} < {\nu _{\min }} \le \nu  \le {\nu _{\max }} < f_{D_{\max}}
\end{IEEEeqnarray}
where ${\nu _{\max }}$ and ${\nu _{\min }}$ denote the maximum and minimum Doppler frequencies of the DPSD, defined as:
\begin{IEEEeqnarray}{rcl}\label{eq:Doppler_range2_1}
{\nu _{\max }} &=& \mathop {\max }\limits_{v \in \left\{ {1,8} \right\}} \left( {{f_{{T_{\max }}}}\cos {\alpha _{{C_v}}} + {f_{{R_{\max }}}}\cos {\beta _{{C_v}}}} \right),\\
\label{eq:Doppler_range2_2}
{\nu _{\min }} &=& \mathop {\min }\limits_{v \in \left\{ {4,5} \right\}} \left( {{f_{{T_{\max }}}}\cos {\alpha _{{C_v}}} + {f_{{R_{\max }}}}\cos {\beta _{{C_v}}}} \right).
\end{IEEEeqnarray}
Hence, the Doppler spread of $S_{hh}\left(\nu\right)$, i.e., $B_d$, is bounded by 
\begin{IEEEeqnarray}{rcl}\label{eq:DS}
B_d=\nu_{\max}-\nu_{\min} \le2f_{D_{\max}},
\end{IEEEeqnarray}
where $2f_{D_{\max}}$ is the Doppler spread, predicted by the conventional models. The right-hand side equality is achievable iff $l_i \to \infty, \forall i$. In this paper, we refer to this feature, described in (\ref{eq:Doppler_range1}--\ref{eq:DS}) as {\it  ``spectral shrinkage.''} In Section IV, this feature will be further discussed with the measured DPSD of \cite{Che13}. \FBB{Note that our discussions on (\ref{eq:Doppler_range1}--\ref{eq:DS}) are limited to the case of stationary RSSs, which are of our main interest in this paper. We are aware that, in V2V channels, mobile scatterers can produce the Doppler shift in the range of $|\nu| \le 4f_{T_{\max}}$ (if the velocities of the Tx, Rx, and mobile scatterers are the same). Analyzing the impact of mobile scatterers on DPSD or Doppler shifts is also an important research topic, which is not in the scope of this research, and hence, left as future work.}  
%
%
%
%

\subsection{Statistical measures}
The MDS $B_1$ and RDS $B_2$ of a DPSD, defined in eq. (3.28) of \cite{Patbook11}, statistically quantify the degree of Doppler spread, so are important measures for the fading channels' rapidity analysis \cite{Ber14, Mol99} and robust receiver design \cite{Mol99,Aco07_2,Cai03}. Therefore, the two quantities will be used in the analysis of our RSS model and model parameter estimation from measurement data in Sections IV and V, respectively. 
When computing the MDS and RDS of the DPSD in (\ref{eq:PSD_total}), a special care is needed due to the mixture of both deterministic LoS and random RSS components. In this case, based on eq. (3.28) of \cite{Patbook11}, the MDS of (\ref{eq:PSD_total}) is obtained as: 
\begin{IEEEeqnarray}{rcl}\label{eq:Mean_Doppler}\nonumber
B_1^{} &=& \int_{-\infty}^{\infty} {\nu \left\{ {\frac{K}{{K + 1}}S_{hh}^{{\rm{LoS}}}\left( \nu  \right) + \frac{1}{{K + 1}}S_{hh}^{{\rm{RSS}}}\left( \nu  \right)} \right\}d\nu } \\
& =& \frac{K}{{K + 1}}{f^{{\rm{LoS}}}} + \frac{1}{{K + 1}}\int_{-\infty}^{\infty} {\nu S_{hh}^{{\rm{RSS}}}\left( \nu  \right)d\nu }.
\end{IEEEeqnarray}
The RDS of (\ref{eq:PSD_total}) can be obtained, similarly as below:
\begin{IEEEeqnarray}{rcl}\label{eq:RMS_Doppler}
B_{2} = \sqrt {\frac{{K{{\left( {{f^{{\rm{LoS}}}} - {B_1}} \right)}^2} + \int_{ - \infty }^\infty  {{{\left( {\nu  - {B_1}} \right)}^2}S_{hh}^{{\rm{RSS}}}\left( \nu  \right)d\nu } }}{{K + 1}}}.
\end{IEEEeqnarray}
As can be seen in (\ref{eq:Mean_Doppler}) and (\ref{eq:RMS_Doppler}), the MDS and RDS are expressed, respectively, as a sum of weighted LoS and RSS components. When a LoS path exists in a V2V channel, both LoS Doppler frequency ${f^{{\rm{LoS}}}}$ and Rician $K$ factor play decisive roles in the determination of MDS and RDS values, and hence have to be included into the analysis.  
It is worth nothing that, for the non-LoS (NLoS) case, the MDS and RDS of (\ref{Proposed_PSD}) become the mean ${m_{{F_D}}}$ and standard deviation $\sigma _{{F_D}}$ of the DPDF, $f_{F_D}\left(\nu\right)$. 
\begin{table*}[!t]
\renewcommand{\arraystretch}{0.9} 
\caption{Model parameters used in Figs. 3--9.}
\label{table1}
\centering
\tabcolsep=0.015cm
\tabcolsep=0.005cm
\footnotesize
\begin{tabular}{|c||c|c|c|c|c|c|}
\hline
\bfseries Param. [unit]& {\bfseries Fig. 3,4} & \bfseries Fig. 5 & {\bfseries Fig. 6} & \bfseries Fig. 7 & \bfseries Fig. 8  & \bfseries Fig. 9\\
\hline
\multicolumn{7}{|c|}{Common parameters: $f_c=5.9$GHz, $\lambda=0.0508$m, $\FB{g_n}=1/{\sqrt N}$}\\
%
\hline
$v_{T}, v_{R}$ [km/h]   &  \multicolumn{3}{c|}{\FB{$105, 105$}} & $87.12, 88.92$ & $105,105$ &   $32.8, 38$\\
\hline
$\gamma_{T}, \gamma_{R}$ [rad.]& \multicolumn{3}{c|}{\FB{in text}}         & $0,0$  &  $0,0$  & $0,\FB{\pi}$  \\
\hline
$x_{T}, y_{T}$ [m]            & \multicolumn{2}{c|}{\FB{$-200, -8.75$}}  & {$\FB{-200}, -5.25$}   &$-30.9, 0$                       &   $-200, -8.75$                 & $-50, -1.75$               \\
\hline
$x_{R}, y_{R}$ [m]           & \multicolumn{2}{c|}{\FB{$200, -8.75$}}   & {$200, -1.75$}  & $30, 0$                            &   $200, -8.75$                 & $50, 1.75$               \\
\hline
$a_1,b_1,c_1,$       & \multicolumn{1}{c|}{\FB{$-263.917, 276.045, 18.364,$}} & \FB{same}  &  \multirow{3}{*}{in text}  & $-49, 46, 14,$    & $-263.917, 276.045, 18.364,$   & $-58.557, 58.753, 8.000,$\\
$d_1, a_2,b_2,$      & \multicolumn{1}{c|}{\FB{$106.396, -263.146, 277.483,$}}  &\FB{as in}&              & $17,-49, 46, $    & $26.396, -263.146, 277.483,$   & $ 13.351, -58.658, 57.919,$ \\
$c_2,d_2$  [m]        & \multicolumn{1}{c|}{\FB{$-103.747, -20.605$}}       &\FB{Fig. 8}     &              & $-17, -14$          & $-23.747, -20.605$                 & $-19.114, -8.003$ \\
\hline
$K$ factor                        & -                                          & $0$                               & $0$                  & $1.175$                                   &  $1.535$                                         & $0.000$ \\
\hline
\end{tabular}
\end{table*}
%

\section{Numerical Results}
In this section, our analytic results of the AoD-AoA PDF in (\ref{eq:JAoDAoA2}), Doppler-AoA PDF in (\ref{eq:joint_Doppler_AoA_PDF}), and DPDF in (\ref{eq:Doppler_PDF}) are validated by histograms. The DPSD-DPDF equivalence in (\ref{eq:PSD_PDF_EQ}) is also validated by using simulations. Based on the justifications made on our analytic results, impacts of RSS layouts on the DPSD shape, Doppler spread, MDS, and RDS are investigated and also compared to  the measured and the modeled DPSDs in \cite{Che13}. Model parameters used in the numerical results are listed in Table \ref{table1}. 

\begin{figure}[t]
    \centering
    \includegraphics [width=8.7cm] {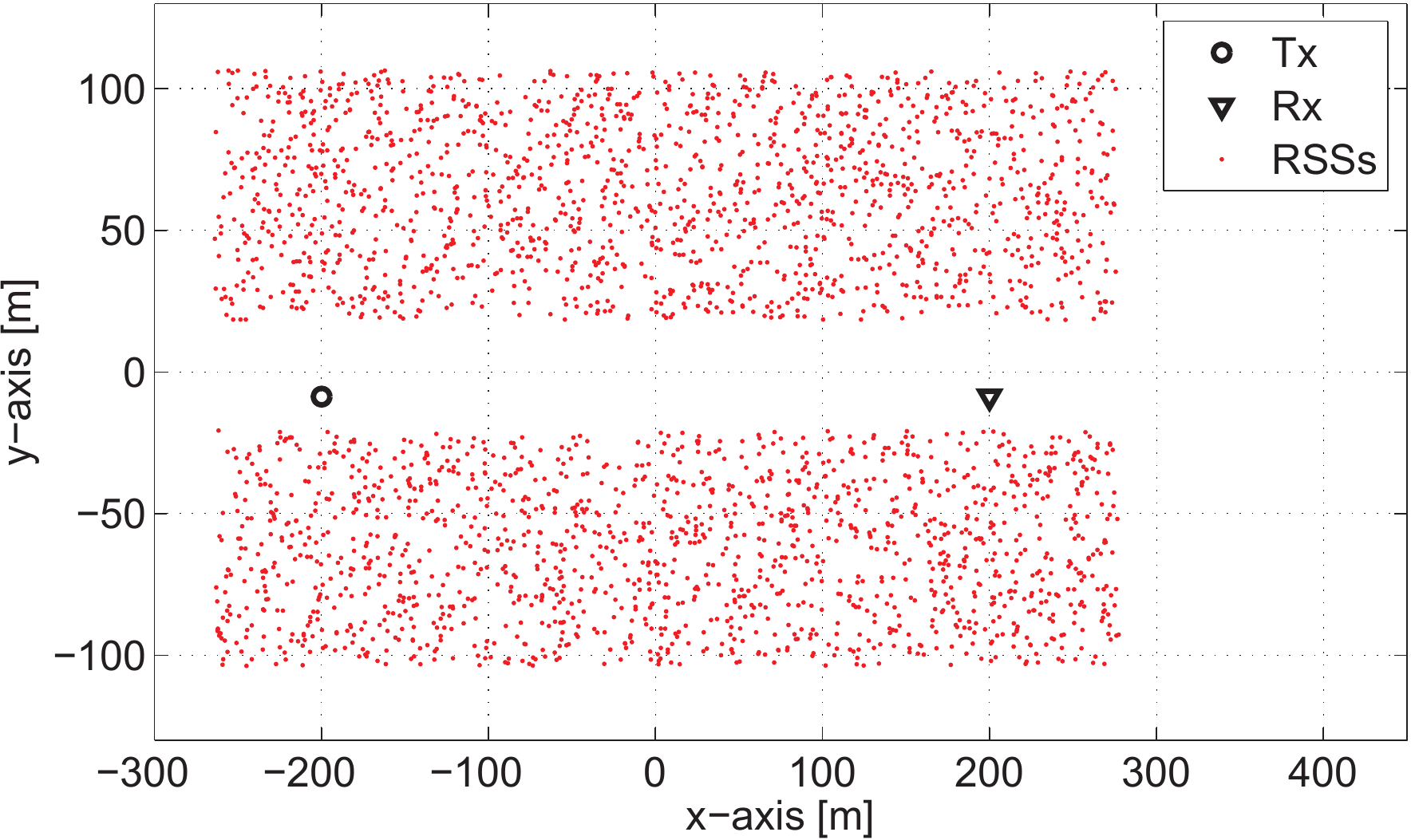}
    \caption{\FB{A scattering plot of the RSSs generated with $N=3000$. The model parameters defining the RSS region ${\cal B}$ are the same as in Fig. \ref{hist_pdf_comp}.}}
    \label{RSS_generation}
\vspace{-0.5cm}
\end{figure}

\subsection{Joint AoD-AoA PDF and Doppler-AoA PDF}
In this subsection, the analytic expressions of the joint AoD-AoA PDF ${f_{{\rm A},{\rm B}}}\left( {\alpha ,\beta } \right)$ in (\ref{eq:JAoDAoA2}) and the joint Doppler-AoA PDF ${f_{{F_D},{\rm B}}}\left( {\nu ,\beta } \right)$ in (\ref{eq:joint_Doppler_AoA_PDF}) are compared with their corresponding normalized histograms. \FB{Most of the model parameters were chosen based on the numerical optimization result from the measurement data set, ``MTM-Expressway Same Direction With Wall, 300-400m,'' as described in Section V. 
Yet, different values were chosen for $d_1$ and $c_2$, to improve the presentation clarity of the joint PDFs and histograms}\footnote{\FB{We chose the $d_1$ ($c_2$) value larger (smaller) than the actual value estimated from the measurement. Otherwise, the widths of the URS and LRS regions become narrower (i.e.,  $d_1 - c_1$ and $d_2 - c_2$ become smaller), and this makes the joint PDF and histogram plots in Fig. 4 difficult to see and interpret. This is due the fact that for narrow widths of the URS and LRS regions, the joint AoD-AoA and Doppler-AoA PDFs have extremely narrow support sets for an independent variable for a given value of the other. }}. For the joint Doppler-AoA PDF,  both SD ($\gamma_T=\gamma_R=0$) and OD ($\gamma_T=0, \gamma_R=\pi$) scenarios \FB{were} considered. 

To obtain the AoD-AoA and Doppler-AoA histograms, in total $N=N_{1}+N_{2}=\FB{10^8}$ RSSs\footnote{To guarantee the equal scattering density, $N_1 = \left\lfloor {N \cdot {A_1}/A} \right\rfloor$ and $N_{2}=N-N_{1}$ numbers of RSSs \FB{were} generated in ${\cal B}_1$ and ${\cal B}_2$, respectively.} \FB{were} randomly generated according to the PDF in (\ref{eq:XY_JPDF}) and then non-linear transformed through (\ref{eq:AoD}), (\ref{eq:AoA}), and (\ref{eq:DF}). \FB{Each histogram was estimated by averaging 100 independent histograms, generated with the total number of bins $M_T$. For clearer presentation, zero bins were excluded in the histogram plots.} To support readers understand, a random scattering plot with $N=3000$ is given in Fig. \ref{RSS_generation}.  

\begin{figure*}[tb]
\centering
\includegraphics[width=\textwidth]{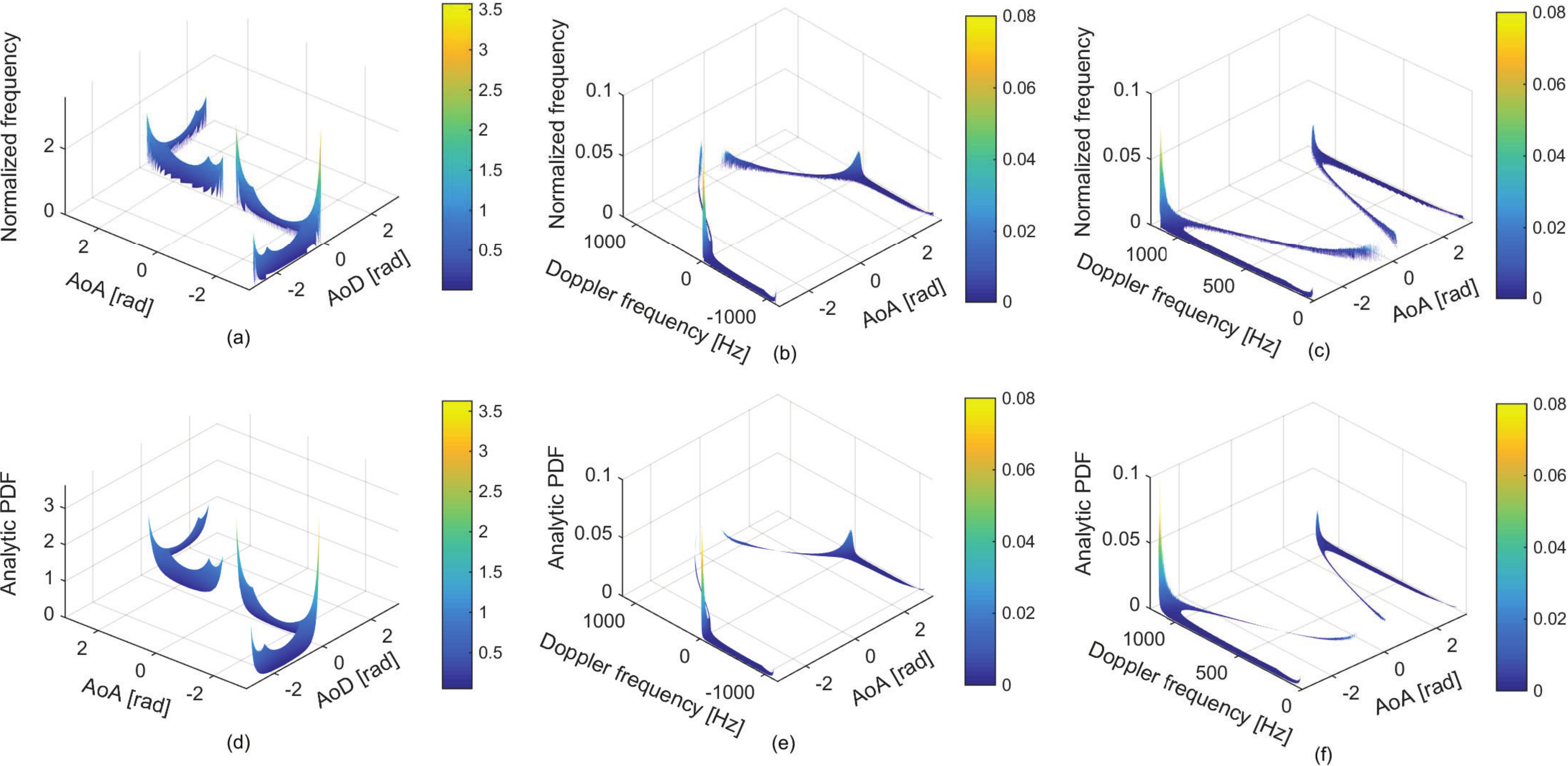}
    \caption{\FB{Comparisons between normalized bivariate histograms and the corresponding analytic joint PDFs. The model parameters used for this figure are listed in Table 1. (a) AoD-AoA histogram; (b) Doppler-AoA histogram for the SD scenario; (c) Doppler-AoA histogram for the OD scenario. In (d)--(f), the respective analytic joint PDFs are presented.}}
    \label{hist_pdf_comp}
\vspace{-0.5cm}
\end{figure*}

\FB{The results in Figs. \ref{hist_pdf_comp}(a)--(f) show that  the two joint PDFs are visually close to their respective normalized histograms. To test the hypothesis that the analytic PDFs closely approximate their respective histogram estimates, we used chi-squared goodness-of-fit test \cite{Ped00, Gub06}. The test statistic $Z$ were computed based on the number of non-empty bins in the histograms, defined as $M$ (note that $M \le M_T$). For the two histograms, their $Z$ values are chi-square distributed with $M-1$ degree of freedom. The $p$ value, defined as $z_p=P(Z>z_{\alpha})$ was chosen to be $0.05$, where $z_{\alpha}$ is the significance level. If $Z \le z_{\alpha}$, we accept the hypothesis; otherwise we reject it. In Table \ref{Chitest_table}, the test results are summarized. Since $Z\le z_{\alpha}$ is satisfied for all histograms, we accept the hypothesis, implying that the two PDFs are good fit to the bivariate histograms. 
}
%
\begin{table*}[t!]
\renewcommand{\arraystretch}{1} 
\caption{\FB{Chi-square test results for the joint AoD-AoA, joint Doppler-AoA, and Doppler frequency PDFs.}}
\label{Chitest_table}
\centering
\tabcolsep=0.2cm
\footnotesize
\begin{tabular}{|c||c|c|c|c|c|}
\hline
\multirow{2}{*}{\bfseries PDFs} & Moving & \multirow{2}{*}{$M_T$} & \multirow{2}{*}{$M$} & \multirow{2}{*}{$Z$} & \multirow{2}{*}{$z_{\alpha}$} \\
 & scenario& & & &  \\
 \hline
Joint AoD-AoA&  - & $9\times10^{6}$ & $551892$ & $8.9\times10^4$ & $55.4\times10^4$\\
\hline
\multirow{2}{*}{Joint Doppler-AoA} & SD & $18\times10^{6}$ & $714218$ & $3.9\times10^{5}$ & $ 7.2\times10^{5}$ \\
\cline{2-6}
                                                                            & OD & $36\times10^{6}$ & $2857031$ & $2.3\times10^{6}$ 
& $ 2.9\times10^{6}$ \\
\hline
\multirow{2}{*}{Doppler frequency} & SD & $8133$ & $8133$ & $87.2$ & $8341.9$\\
\cline{2-6}
											     & OD & $4067$ & $4067$ & $967.7$ & $4214.4$\\
\hline
\end{tabular}
\end{table*}


The joint AoD-AoA PDF in Fig. \ref{hist_pdf_comp}d shows high dependency between AoD and AoA. This is the consequence of SB scattering, where a location of a RSS uniquely determines a pair of AoD and AoA. Such statistical dependency can be also found in the simulated bi-azimuth power spectrum result in \cite{Zho12}, where its domain shape is similar to the result in Fig. \ref{hist_pdf_comp}d. These results clearly demonstrate that the independence assumption between AoD and AoA in \cite{Ava11} is not suitable for SB scattering models. 

\FB{The 2D placement of RSSs and the SB scattering mechanism make the shape of the joint AoD-AoA PDF distinctive. In particular, intermediate density values appear in ${\cal A}_k$ in (\ref{eq:Samplespace1}) for $k=1,4,5,8$ (refer to Fig. \ref{Subspace}). This corresponds to the case when $\left(\alpha,\beta\right)$ is close to $(0,0)$, $(\pi,\pi)$, and $(-\pi,-\pi)$. Meanwhile, high density values appear near $(0,\pi)$ and $(0,-\pi)$. 
Such intermediate/high density values in the joint AoD-AoA angles lead to high density values in the joint Doppler-AoA PDF around specific Doppler frequencies. For the SD and OD scenarios, these frequencies can be obtained by substituting the angles into (\ref{eq:DF}) with proper moving directions. In Table \ref{table:DF}, the Doppler frequency $F_D$ for those AoD-AoA pairs are summarized. From the results in Table \ref{table:DF} and our discussion above, it is easy to anticipate that the joint Doppler-AoA PDF will have high density values near the maximum, minimum, and relative Doppler frequencies, i.e., $f_{D_{\max}}$, $-f_{D_{\max}}$, and $\nu_{\rm rel}=f_{T_{\max}}-f_{R_{\max}}$ ($0$Hz in this case), respectively, for the SD scenario. In the OD scenario, high density values may appear near $\nu_{\rm rel}=-\nu_{\rm rel}=0$Hz and $f_{D_{\max}}$ in the joint Doppler-AoA PDF. In fact, these observations agree with our simulation and numerical analysis shown in Figs. 4(b), (c), (e), and (f).} 
\begin{table*}[!t]
\renewcommand{\arraystretch}{1} 
\caption{\FB{Summary of the Doppler frequencies $F_D$, corresponding to the AoD-AoA pairs, $(\alpha, \beta)$, at which ${f_{{\rm A},{\rm B}}}\left( {\alpha ,\beta } \right)$ have high density values.}}
\label{table:DF}
\centering
\tabcolsep=0.2cm
\footnotesize
\begin{tabular}{|c|c|c|}\hline
\multirow{2}{*}{ $\left( {\alpha ,\beta } \right)$}& \multicolumn{2}{c|}{Doppler frequency $F_{D}$  in (14)}\\
\cline{2-3}
                     & SD scenario $\left( {\gamma_T = \gamma_R=0} \right)$ & OD scenario $\left( {\gamma_T =0, \gamma_R=\pi} \right)$\\
\hline
$\left( {0,0 } \right)$ & $f_{T_{\max}} + f_{R_{\max}}=f_{D_{\max}}$ & $f_{T_{\max}} - f_{R_{\max}}=\nu_{\rm rel}$ \\
\hline
$\left( \pi,\pi \right)$,$\left(-\pi,-\pi \right)$   &$ - f_{T_{\max}} - f_{R_{\max}}=-f_{D_{\max}}$ & $-f_{T_{\max}} + f_{R_{\max}}=-\nu_{\rm rel}$\\
\hline
$\left( {0,\pi } \right)$,$\left( {0,-\pi } \right)$  & $f_{T_{\max}} - f_{R_{\max}}=\nu_{\rm rel}$ & $f_{T_{\max}} + f_{R_{\max}}=f_{D_{\max}}$\\
%
\hline
\end{tabular}
\end{table*}
\subsection{DPDF and DPSD}
\begin{figure}[t]
    \centering
    \includegraphics [width=8.7cm] {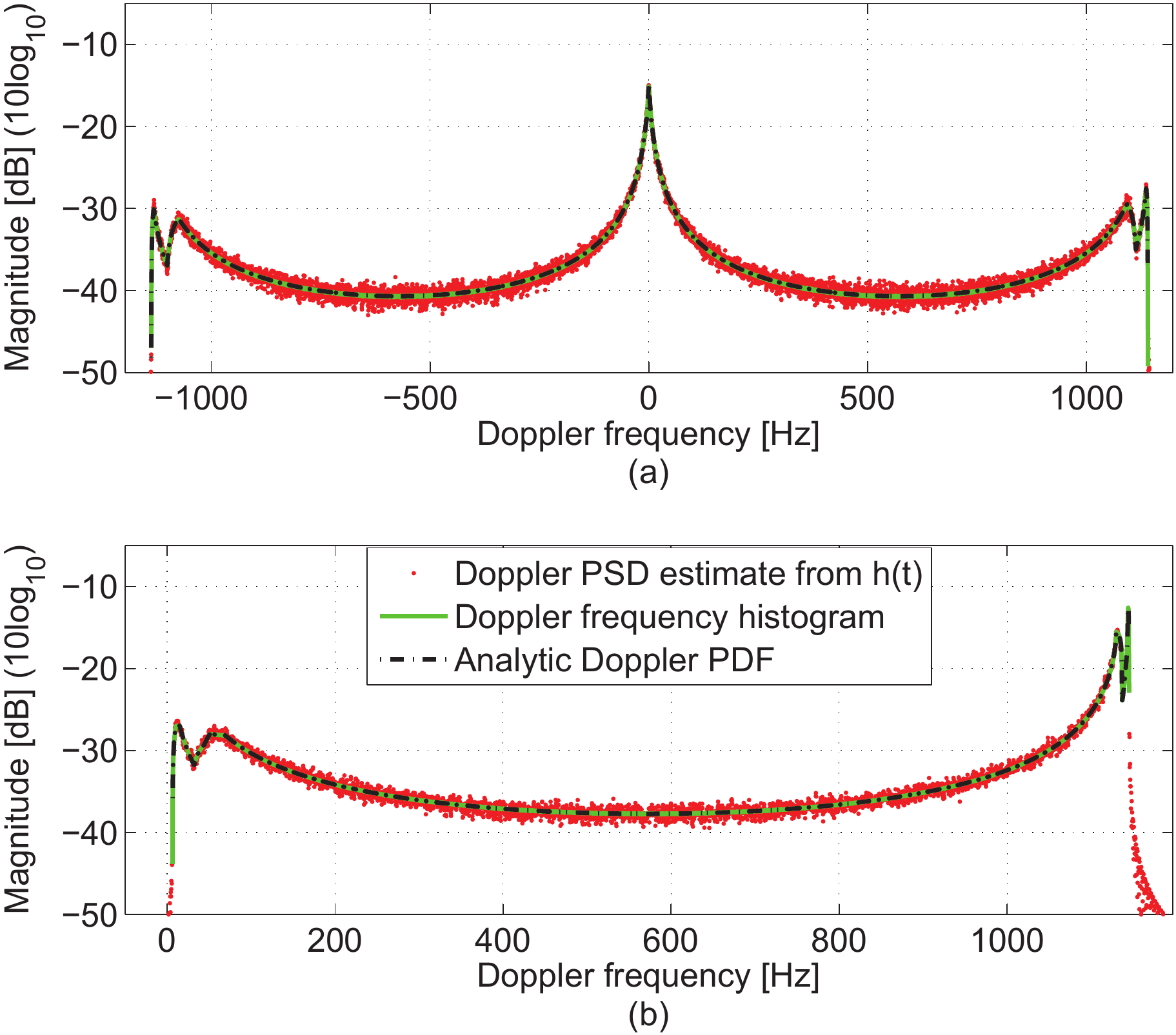}
    \caption{\FB{A comparison between the DPSD estimate from $h\left(t\right)$ in (\ref{eq:TotalCG}), the normalized Doppler frequency histogram, and the DPDF $f_{F_D}\left(\nu\right)$ in (\ref{eq:Doppler_PDF}) for (a) SD scenario and  (b) OD scenario.}}
    \label{Doppler_PSD_comp1}
\vspace{-0.5cm}
\end{figure}

This subsection aims 1) to show the validity of the analytic DPDF $f_{F_D}\left(\nu\right)$ in (\ref{eq:Doppler_PDF}); 2) to experimentally validate the equivalence in  (\ref{eq:PSD_PDF_EQ}); and \FB{3) to explain the characteristics of the DPSD.} 
For this purpose, in Fig. \ref{Doppler_PSD_comp1}, the analytic $f_{F_D}\left(\nu\right)$, a normalized Doppler frequency histogram $\hat{f}_{F_D}(\nu)$, and a DPSD estimate of $h(t)$, denoted as $\hat{S}_{hh}(\nu)$, are compared for SD and OD scenarios.
\FB{The model parameters were chosen based on the numerical optimization result in Section V-B. $K=0$ was chosen as our interest is the DPSD and DPDF of the diffuse component. Note that $\hat{f}_{F_D}(\nu)$ was estimated by averaging 100 times of independent histograms, which are generated with \FB{$N=5\times10^6$}.} To obtain $\hat{S}_{hh}(f)$, we at first generated discrete-time channel gains $h[k]$ for $t\in[0, 2]$sec with \FB{$N=10^4$}, according to  (\ref{eq:RSS_gain}). The sampling frequency was $f_s = \FB{8}f_{D_{\rm max}}$. Then we computed ACF and averaged it over \FB{$200$} times. Finally, the fast Fourier transform was taken to obtain $\hat{S}_{hh}(\nu)$.

The results in Fig. \ref{Doppler_PSD_comp1} show that $f_{F_D}\left(\nu\right)$ in (\ref{eq:Doppler_PDF}) is close not only to the normalized histogram but also to the DPSD estimate. \FB{The chi-square test result quantitatively supports the close agreement between the histogram and DPDF, see Table \ref{Chitest_table}. Note that the mean square error (MSE) between $f_{F_D}\left(\nu\right)$ and $\hat{S}_{hh}(\nu)$ are $3.5\times 10^{-4}$, which is fairly small and also comparable to the MSE between $f_{F_D}\left(\nu\right)$ and $\hat{f}_{F_D}(\nu)$ (i.e., $2.6\times 10^{-4}$). These results clearly validate (\ref{eq:Doppler_PDF}) and (\ref{eq:PSD_PDF_EQ}).}

In Fig. \ref{Doppler_PSD_comp1}(a), the DPSD shows an {\it ``incomplete W-shape,''} where two weak peaks and a single strong peak appear around $\nu_{\max}\approx\FB{1140}$Hz, $\nu_{\min}\approx-\FB{1137}$Hz, and $\nu_{\rm rel}=0$Hz in the SD scenario, respectively. This spectral tendency coincides with various SD measurement results in V2V channels  \cite{Che13,Tan08, Aco07,Aco07_2}. In Fig. \ref{Doppler_PSD_comp1}(b), the DPSD in the OD scenario shows an {\it ``incomplete U-shape,''} where weak and strong peaks appear near $\nu_{\min}\approx\FB{6}$Hz and $\nu_{\max}\approx\FB{1145}$Hz, respectively. Note that such a spectral shape can be found in the DPSDs measured in the expressway and urban canyon oncoming (or OD) scenarios in \cite{Aco07, Aco07_2}.

\FB{Such spectral peaks of the DPSD are due to the 2D placement of RSSs, parallel to the moving directions of the Tx and Rx, and SB scattering. As a result, scattered signals propagate via specific joint angles with high probabilities, as summarized in Table \ref{table:DF}, and thereby resulting in spectral peaks around the specific Doppler frequencies given below:
\begin{enumerate}
\item SD scenario: $f_{D_{\max}}$, $-f_{D_{\max}}$, and $\nu_{\rm rel}$.
\item OD scenario: $\nu_{\rm rel}$, $-\nu_{\rm rel}$, and $f_{D_{\max}}$.
\end{enumerate}
For the the model parameters used in Fig. \ref{Doppler_PSD_comp1}, it becomes $f_{T_{\max}}=f_{R_{\max}}\approx573.6$Hz, $f_{D_{\max}}\approx1147.0$Hz, and $\nu_{\rm rel}$=0Hz. Consequently, the DPSD has three (two) spectral peaks near those frequencies for the SD (OD) scenario.} 



\subsection{Impacts of RSS layouts on the DPSD, Doppler Spread, MDS, and RDS}

The 1D RSS model in \cite{Che13} assume that RSSs are distributed in two lines having infinite length. In reality, however, it is nearly impossible to receive the signals coming from RSSs at infinite distances due to geographical limits (such as curves and hills) and path-loss, and hence it is more practical to model the RSSs to be distributed in finite areas as in Fig. \ref{RSS_model}. According to the DPSD measurement results in \cite{Che13}, limitations in length lead to the measured DPSD, having narrower Doppler spread by $10$-$15\%$ than the theoretically expected one. In addition, it was also observed in \cite{Che13} that the degree of the spectral shrinkage varies depending on the width of the unobstructed area in the measurement environments\footnote{In \cite{Che13}, the DPSD measured in the \FB{rural} area has smaller spectral shrinkage than those measured in the highway. Note that the road widths in the \FB{rural} and highway environments were $23$m and $60$m, respectively. }. 
Hence, this subsection is devoted to clarify how the layout of the RSS region can impact on channel Doppler characteristics, i.e., DPSD shape, Doppler spread \FB{$B_d$}, MDS $B_1$, and RDS $B_2$. 

\begin{figure*}[tb]
\centering
\includegraphics[width=\textwidth]{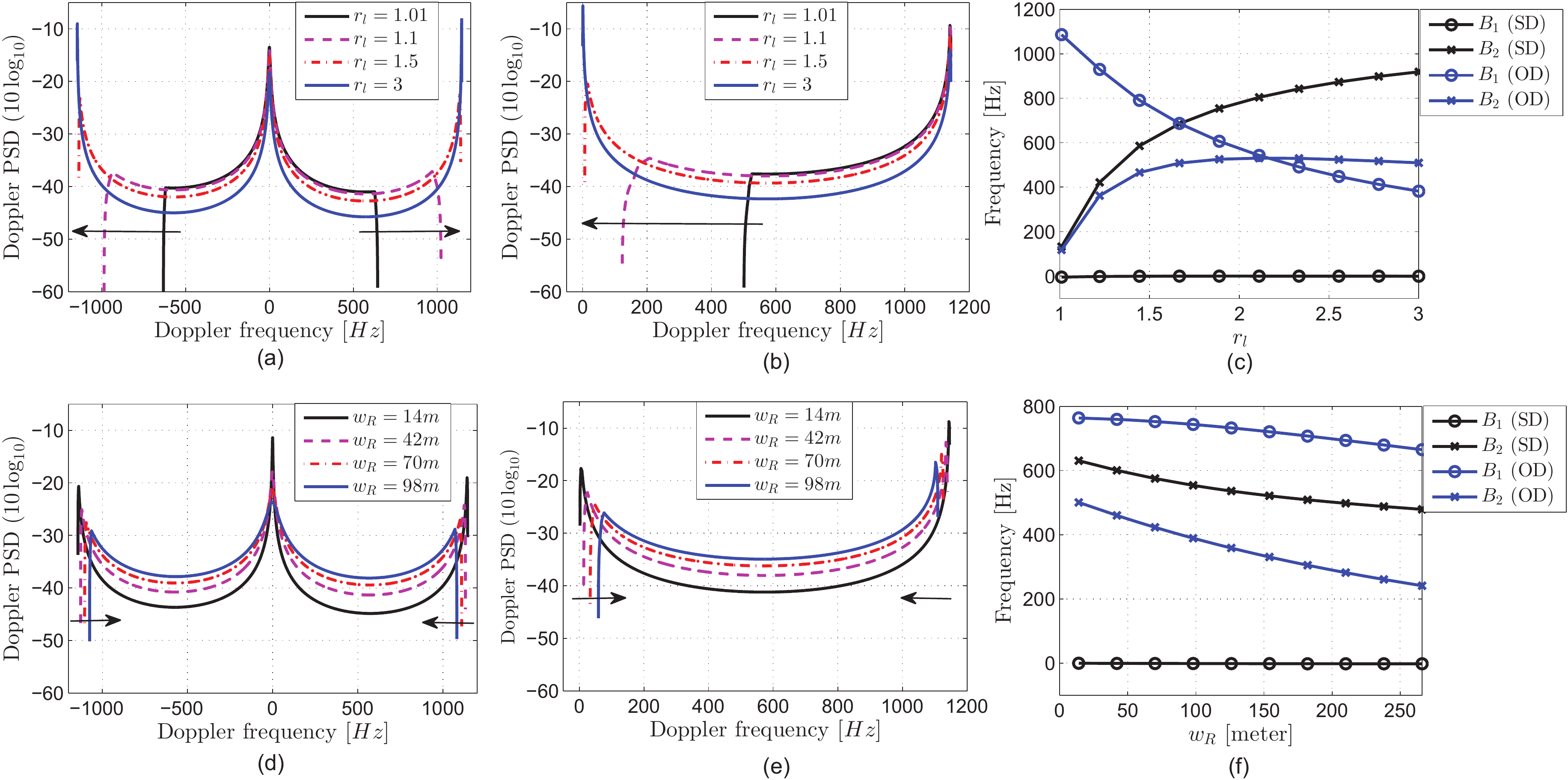}
    \caption{Analysis of the DPSD $S_{hh}\left(\nu\right)$ in (\ref{Proposed_PSD}) and the corresponding MDS $B_1$ and RDS  $B_2$ for different $r_l$ and $w_R$ values. (a) $S_{hh}\left(\nu\right)$ for different  $r_l$ values with $w_R=28$m in the SD scenario. (b) $S_{hh}\left(\nu\right)$ for the OD scenario with the same parameters as in (a). (c) $B_1$ and $B_2$ w.r.t. $r_l$ for the SD and OD scenarios ($w_R=28$m). (d)--(f) are the respective counterparts of (a)--(c), but with different road width $w_R$ at a fixed ratio $r_l\approx1.50$.}
    \label{DopplerPSD_diff_length_and_width}
\end{figure*}

In Fig. \ref{DopplerPSD_diff_length_and_width}, the DPSD $S_{hh}\left(\nu\right)$ for $\nu\in[\nu_{\min}, \nu_{\max}]$ in (\ref{Proposed_PSD}) as well as its MDS $B_1$ and RDS $B_2$ are analyzed for the ratio of the length of the RSS region to the LoS distance, i.e., $r_{l} = l/d_{\rm LoS}$ ($l=l_1=l_2$), and the road width $w_R$. To this end, the following model parameters were used in the numerical analysis: \FB{$a_{i}=-0.5d_{\rm LoS}r_{l}$, $b_i=0.5d_{\rm LoS}r_{l}$}, $c_1=0.5w_{R}$, $d_1 = c_1+5$, $c_2=d_2-5$, $d_2 = -0.5w_{R}$ meters for $i\in\{1,2\}$\footnote{Note that this parameterization was used to keep the RSS region symmetric to $x$-axis and $y$-axis for the simplification of the resulting DPSD shape based on the given $r_l$ and $w_R$ values.}. \FB{The vehicle location parameters, i.e., $x_T, y_T, x_R, y_R$, were chosen based on 
the assumption on the lane width $3.5$m\footnote{Note that 2.7 to 3.6m lane width are used in general U.S. roads, where 3.6m width is typical for most of the U.S. highways \cite{GB11}. In this paper, 3.5m lane width is assumed for all numerical results for consistency.} and the LoS distance $d_{\rm LoS}\approx400$m, maintained during the measurement in the expressway SD environment \cite{Aco07_2}.} The rest parameters are listed in Table 1. 

In Fig. \ref{DopplerPSD_diff_length_and_width}a, the DPSD was analyzed for different values of $r_l$ with $w_R=28$m (SD scenario). The result shows that as the length of the RSS region, relatively to the LoS distance, increases, 1) the Doppler spread $B_d$ increases from about $f_{D_{\max}}$ to $2f_{D_{\max}}$; 2) the DPSD values at $\nu_{\min}$ and $\nu_{\max}$ increase while decreasing at $\nu_{\rm rel}$ and elsewhere; and 3) the overall spectrum shape changes from {\it incomplete to complete W-shape}. Similar observations for the OD scenario can be found in Fig. \ref{DopplerPSD_diff_length_and_width}b. Yet, $B_{d}$ riches up to $f_{D_{\max}}$, and the spectrum changes from {\it incomplete to complete U-shape}. These spectral changes also lead to the variations in channel statistical properties. Fig. \ref{DopplerPSD_diff_length_and_width}c shows the MDS $B_1$ and the RDS $B_2$ w.r.t. $r_l$. It was shown that both quantities dramatically changes as the length of the RSS region varies. Note that $B_1$ values for the SD scenario stay near 0Hz due to the symmetric placements of RSSs and $v_{T}=v_{R}$. 

Figs. \ref{DopplerPSD_diff_length_and_width}d--f are the counterparts of Figs. \ref{DopplerPSD_diff_length_and_width}a--c with a fixed $r_l\approx1.50$ but for different values of the road width $w_R$. Figs. \ref{DopplerPSD_diff_length_and_width}d and f show that as the road width increases, 1) $B_d$ decreases from about $2f_{D_{\max}}$ to a smaller quantity; and 2) the DPSD shape, in general, becomes flattened out. Finally, Fig. \ref{DopplerPSD_diff_length_and_width}f demonstrates the changes in the channel statistical properties as the road width increases. For the SD scenario, $B_1\approx 0$ for all $w_R$ due to the same reason as in Fig \ref{DopplerPSD_diff_length_and_width}c. 

From the above observations, it is apparent that the ratio $r_l$ and the road width $w_R$ have critical impacts on the DPSD shape, Doppler spread, MDS, and RDS. \FB{The most important factor that makes the DPSD shape change from the incomplete to complete W/U-shapes is the length of the RSS region $l$, relative to the LoS path distance $d_{\rm LoS}$. As $l$ becomes larger than $d_{\rm LoS}$, more signals come from the pairs of AoD and AoA close to $(0,0)$, $(\pi,\pi)$, and $(-\pi,-\pi)$, thereby increasing the probability density around those angles in the joint AoD-AoA PDF. For the SD scenario, these angles correspond to $f_{D_{\max}}$, $-f_{D_{\max}}$ (See Table \ref{table:DF}). Accordingly, the outmost DPSD values become increase and the DPSD shape becomes complete W-shape. On the other hand, if $l$ becomes smaller relative to $d_{\rm LoS}$, the range of the AoD/AoA becomes smaller. Also, the probability density near $(0,0)$, $(\pi,\pi)$, and $(-\pi,-\pi)$ becomes smaller. Eventually, the Doppler frequency range, i.e., $\nu\in[\nu_{\min},\nu_{\max}]$,  becomes smaller, and the outmost DPSD values become decrease. In this case, the DPSD shape becomes incomplete W-shape. For the OD case, a similar mechanism applies. Increasing $l$ makes the DPSD value near $\nu_{\rm rel}$ ($0$Hz if $f_{T_{\max}}=f_{R_{\max}}$) increase. Hence, the DPSD becomes complete U-shape. If $l$ decreases, $\nu_{\min}$ increases and the DPSD value at $\nu_{\min}$ becomes decrease. In this case, the DPSD shape becomes incomplete U-shape.}

Compared to 
the 1D RSS model of \cite{Che13}, the RSS model in Fig. \ref{RSS_model} predicts smaller Doppler spread (depending on the RSS layouts), which is more close to the reality. 
\begin{figure}[t]
    \centering
 \includegraphics [width=8.7cm] {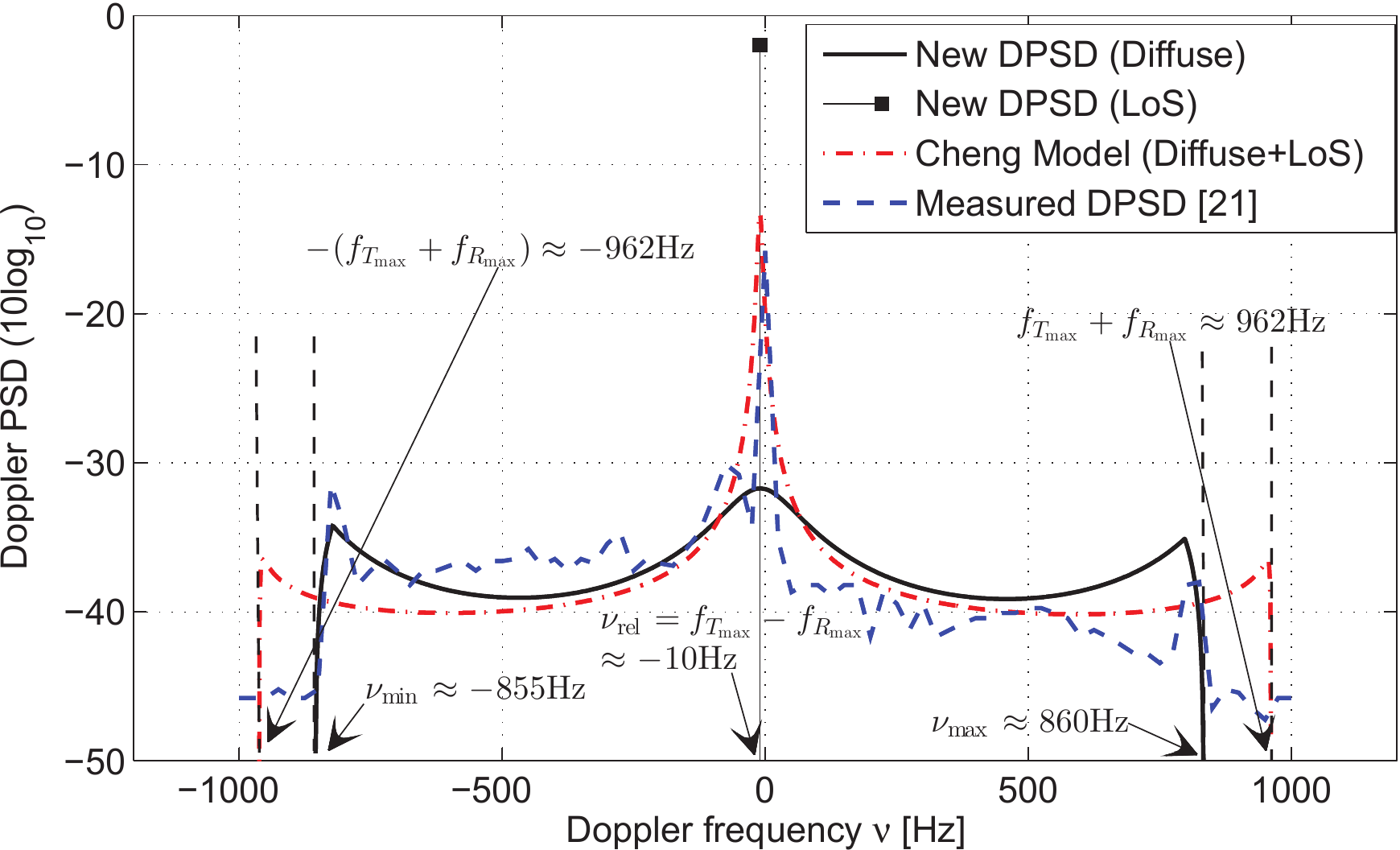}
    \caption{A comparison between the new analytic DPSD in (\ref{Proposed_PSD}), Cheng's 1D RSS model, and the measured DPSD presented in Fig. 9c of \cite{Che13}. The three DPSDs are normalized to the unit area. The approximate values of the maximum, minimum, and relative Doppler frequencies of the new analytic DPSD are also presented. The maximum and minimum possible Doppler frequencies are $962$Hz and $-962$Hz, respectively.}
    \label{MComp1}
\end{figure}
To demonstrate this, Fig. \ref{MComp1} shows a comparison of the new analytic DPSD in (\ref{Proposed_PSD}) with the DPSD of the model in \cite{Che13}, referred as Cheng's model, and the measured DPSD presented in Fig. 9c of \cite{Che13}. The measurement was performed at a carrier frequency of 5.9 GHz in \FB{rural} LoS environments of Pittsburgh, PA. The measurement parameters are: $d_{\rm LoS}=60.9$m, $v_T=24.2$m/s, $v_R=24.7$m/s, $f_{T_{\rm max}}\approx 476$Hz, $f_{R_{\rm max}}\approx 486$Hz, $\gamma_T=\gamma_R=0$ (SD). To reproduce Cheng's model, the same model parameters were used as in \cite{Che13} while a scale parameter was manually chosen to closely approximate the result in Fig. 9c of \cite{Che13}. \FB{For the LoS component, the definition in Appendix of \cite{Che13} was used}. Meanwhile, the new analytic DPSD was generated based on the parameters listed in Table 1 which closely approximate the overall shape, range, and peak positions of the measured DPSD. 
\FB{Note that we chose the LoS powers of the Cheng's and the new DPSDs in such a way that the RDS values of the two models are closely approximated to the measurement counterpart. The corresponding Rician $K$ factors of Cheng's and new DPSDs are $2.580$ and $1.715$, respectively. All three DPSDs in Fig. 7 were normalized to the unit area. To show the goodness of fit of the models, the Doppler spread, MDS, and RDS values of the three DPSDs are summarized in Table \ref{GOFEval}\footnote{\FB{The negative MDS value of the measured DPSD is due to the asymmetric antenna gain during the measurement, as mentioned in \cite{Che13}. Hence, it was not possible to estimate the two models' parameters, making their MDS values further close to the MDS value of the measured DPSD while preserving the RDS accuracy and visual similarities.}}.}

Fig. \ref{MComp1} shows that the measured PSD is in an {\it ``incomplete W-shape,''} where the three peaks appear around $-825$Hz, $-10$Hz, and $825$Hz. The Doppler spread $B_d$ of the measurement data, defined as the width of the frequency interval of which the measured DPSD is above the noise level, is about $1700$Hz. Note that this width is about $12\%$ less than the theoretically expected Doppler spread, i.e., $2f_{D_{\max}}\approx1924$Hz. The central peak has the highest value due to the LoS component while the right hand side of the spectrum values are lower than the other side. This is due to the non-symmetric antenna gain pattern during the measurement \cite{Che13}. Even though this asymmetry, both models well describe the overall spectral features while the new analytic DPSD more precisely captures the positions of the spectral peaks and the range of the measured DPSD \FB{due to the 2D model geometry, bounded in length.} Cheng's model assumes that RSSs are placed on infinite lines, thereby overestimating $B_d$. The result clearly shows the practicality of using the finite 2D RSS model in Fig. \ref{RSS_model}. 

%
\begin{table}[!t]
\renewcommand{\arraystretch}{1.0} 
\renewcommand{\arraystretch}{1.1} 
\caption{\FB{The Doppler spread, MDS, and RDS comparisons between the measured, Cheng's and new DPSDs in Fig. 7.}}
\label{GOFEval}
\centering\
\tabcolsep=0.015cm 
\footnotesize
\begin{tabular}{|c||c|c|c|}
\hline
\bfseries Measure [unit] & Measured DPSD \cite{Che13} & Cheng's DPSD & New DPSD \\
\hline
$B_d $[Hz]  & 1700 & 1924 & 1715 \\
\hline
MDS [Hz] & -58.8 & -8.7 & -15.2 \\
\hline
RDS [Hz] & 269.6 & 269.6 & 269.6\\
\hline
\end{tabular}
\end{table}

\section{Comparisons with Measurement Data}

In this section, we compare the new analytic DPSD, $S_{hh}\left(\nu\right)$ in (\ref{Proposed_PSD}), with the measured DPSD, collected for the channel model development in support of the IEEE 802.11p standard working group \cite{Aco07, Aco07_2}. Among the six different measurement data sets, we selected ``MTM-Expressway Same Direction With Wall, 300-400m'' and ``MTM-Urban Canyon Oncoming 100m'' for comparison of SD and OD scenarios, respectively. The main reason for choosing the datasets is due to detailed descriptions of the measurement setups, locations, data processing procedures, per-tap measured DPSD, and modeled delay-Doppler profiles available in \cite{Aco07_2}. Most importantly, the two measurement data sets are suitable for investigating the impact of RSSs on the DPSD of V2V channels, as they were obtained in various expressways in Atlanta, Georgia, and Edgewood Avenue in Downtown Atlanta, respectively, where sound blockers, dense trees, and buildings are placed along the \FB{straight} roadsides (see Figs. 93 and 96 of \cite{Aco07_2}). Note that the measured DPSD were obtained by averaging over a large number of 0.6s-long segments recorded in a same location or over different similar locations (see Table 2 of \cite{Aco07_2}). Hence, the received power originated from vehicles, quickly moving away from the Tx and Rx, is likely to be averaged out while the DPSD features due to RSSs in regular positions are clearer. 

The two data sets consist of 8 and 5 delay taps, respectively, where each tap has a unique measured DPSD. Since our interest is to compare the analytic DPSD created by the total RSS region in Fig. \ref{RSS_model} with the measured data, we summed the per-tap measured DPSDs over all delay taps  for each data set and then normalized them (unit area), in order to obtain the two ``total measured DPSDs,'' shown in Figs. \ref{MComp2} and \ref{MComp3}, respectively. For convenience, we denote the two measured spectra as $\tilde{S}^{SD}[m]$ and $\tilde{S}^{OD}[m]$. For both spectra, $m\in\{1,2,..., M\}$ denotes the frequency index with the number of measurement samples $M$. The Doppler spread of a measured spectrum is defined as $\tilde{B}_d=\tilde{\nu}_{\max} - \tilde{\nu}_{\min}$, where $\tilde{\nu}_{\max}$ and $\tilde{\nu}_{\min}$ denote the maximum and minimum Doppler frequencies above the noise level. The sampling interval is given by $\Delta\nu=\tilde{B}_d/(M-1)$. Hence, the Doppler frequency of the $m$th measurement sample can be calculated by $\nu_{m}=\tilde{\nu}_{\min}+(m-1)\Delta{\nu}$. The aforementioned parameters for each data set are summarized in Table \ref{table2}. 

It is noteworthy that the powers within $\tilde{S}^{SD}[m]$ and $\tilde{S}^{OD}[m]$ are mostly due to random diffuse and discrete scattering, as the deterministic parts of the measured spectra given in Chapter 7 of \cite{Aco07_2} were removed during the post processing. The optimization problem formulation for the DPSD comparisons, and the corresponding results for each data set will be given in next subsections. 

%
\begin{table}[!t]
\renewcommand{\arraystretch}{1.0} 
\renewcommand{\arraystretch}{0.8} 
\caption{Optimization Parameters and the Corresponding Model Errors for the Two Total Measured DPSDs, obtained from \cite{Aco07_2}.}
\label{table2}
\centering\
\tabcolsep=0.015cm 
\footnotesize
\begin{tabular}{|c|c|c|}
\hline
{\bfseries Measured total} & \multirow{2}{*}{Opt. parameters} & Model errors\\
{\bfseries DPSD   } &                             & (LSE, MSE, MDSE, RDSE\tablefootnote{Each abbreviation represents: LSE-lease square error; MSE-mean square error; MDSE-MDS error; RDSE-RDS error.})\\
\hline
\multirow{5}{*}{$\tilde{S}^{SD}$}  & $\tilde{\nu}_{\min} =-1200$Hz,                        & $1.105 \times 10^{-5}$,\\
                                             &   $\tilde{\nu}_{\max}=1200$Hz,                       & $9.132 \times 10^{-7}, $\\                      
                                             & $\Delta{\nu} = 20$Hz, $M=121$,             & 0.001Hz,\\
                                             & $\varepsilon_1=\varepsilon_2=0.001$Hz.     & 0.001Hz\\
                                             & ${\tilde B}_1\approx8$Hz, ${\tilde B}_2\approx315$Hz  & \\
\hline
\multirow{5}{*}{$\tilde{S}^{OD}$}  & $\tilde{\nu}_{\min} =-880$Hz,                        & $1.074 \times 10^{-3}$,\\
                                             &   $\tilde{\nu}_{\max}=820$Hz,                       & $1.249 \times 10^{-5}$, \\                      
                                             & $\Delta{\nu} = 20$Hz, $M=86$,             & 0.005Hz,\\
                                             & $\varepsilon_1=\varepsilon_2=0.01$Hz.     & 0.010Hz.\\
                                             & ${\tilde B}_1\approx328$Hz, ${\tilde B}_2\approx90$Hz  & \\
\hline
\end{tabular}
\end{table}

\subsection{Optimization problem formulation}

The problem of model comparison to measurement data can be understood as an optimization problem, i.e, finding a best set of model parameters, which minimize the difference (or some error metric) between the model and data. The estimated parameters should not only satisfy the geometrical constraints imposed by the model assumptions, but also be physically reasonable w.r.t. the underlying measurement environment. Note that $f_c$, $v_T$, $v_R$, $\gamma_T$, $\gamma_R$ are given from the measurement set up in \cite{Aco07_2}. The location parameters, i.e., $x_T, y_T, x_R, y_R$, can be arbitrarily chosen based on the LoS distance $d_{\rm LoS}$ maintained during the measurements and the assumption on the lane width $3.5$m. 
The rest of model parameters, expressed in a vector form, ${\bf x}=\left(a_1, b_1, c_1, d_1, a_2, b_2, c_2, d_2, K\right)^{\rm T}$, are needed to be found via numerical optimizations. Note that the Rician $K$ factor was included in ${\bf x}$, to estimate the spectral power, which cannot be explained by only RSSs. 

To estimate ${\bf x}$, we aim to solve the following constrained least square error (LSE) problem defined as below: 
\vspace{-0.8cm}
\begin{IEEEeqnarray}{l}\label{eq:Opt_problem}
{\rm{minimize ~}}\sum\limits_{m = 1}^{M} {\left\{ {{{\tilde S}[m]} - {S}\left( {{\nu _m},{\bf{x}}} \right)} \right\}}^2,\\\label{eq:Const1}
{\rm{~ subject ~to:~}} {\FB{q}_i}\left( {\bf{x}} \right) \le {\varepsilon _i},\\\label{eq:Const2}
~~~~~~~~~~~~~~~~~{\bf{Ax}} \le {\bf{b}},\\\label{eq:Const3}
~~~~~~~~~~~~~~~~~{{\bf{x}}_L} \leqq {\bf{x}} \leqq {{\bf{x}}_U},
\end{IEEEeqnarray}
%
%
where ${\tilde S}{[m]}$ and ${S}\left( {{\nu _m},{\bf{x}}} \right)$ denote the measured and analytic DPSD values at an index $m$, respectively. 
Note that ${\FB{q}_i}\left( {\bf{x}} \right) = \left| {{\tilde{B}_i} - {{B}_i}\left( {\bf{x}} \right)} \right|, i\in\{1,2\}$, where ${\tilde B}_1$ and  ${\tilde B}_2$ are the MDS and RDS of ${\tilde S}{[m]}$, and ${B}_i\left( {\bf{x}} \right)$ for $i\in\{1,2\}$ are the respective counterparts for ${S_{hh}}\left( {{\nu},{\bf{x}}} \right)$. $\varepsilon_1$ and $\varepsilon_2$ are the maximum absolute errors on the MDS and RDS. In (\ref{eq:Const2}), ${{\bf{Ax}} \le {\bf{b}}}$ is designed to upper bound the road width $w_R=c_1-d_2$ depending on the measurement environments and to ensure $d_i - c_i \ge 3$m, $\forall i$. Similarly, the inequalities in (\ref{eq:Const3}) are used to properly limit the range of $\bf x$, according to (\ref{eq:MConst1}) and measurement environments. Note that $\leqq$ denotes an element-wise inequality between two vectors, and ${\bf{x}}_L$ (${\bf{x}}_U$) is a vector, whose elements are lower (upper) bounds on each element of $\bf x$. To find a local minimum $\bf{x}^*$ from (\ref{eq:Opt_problem})--(\ref{eq:Const3}), an Active-set algorithm was used. The results of the DPSD comparisons will be given in the following subsections.

\subsection{Model comparison with the data set, ``MTM - Expressway Same Direction With Wall, 300-400m''}
\begin{figure}[t]
    \centering
    \includegraphics [width=8.7cm] {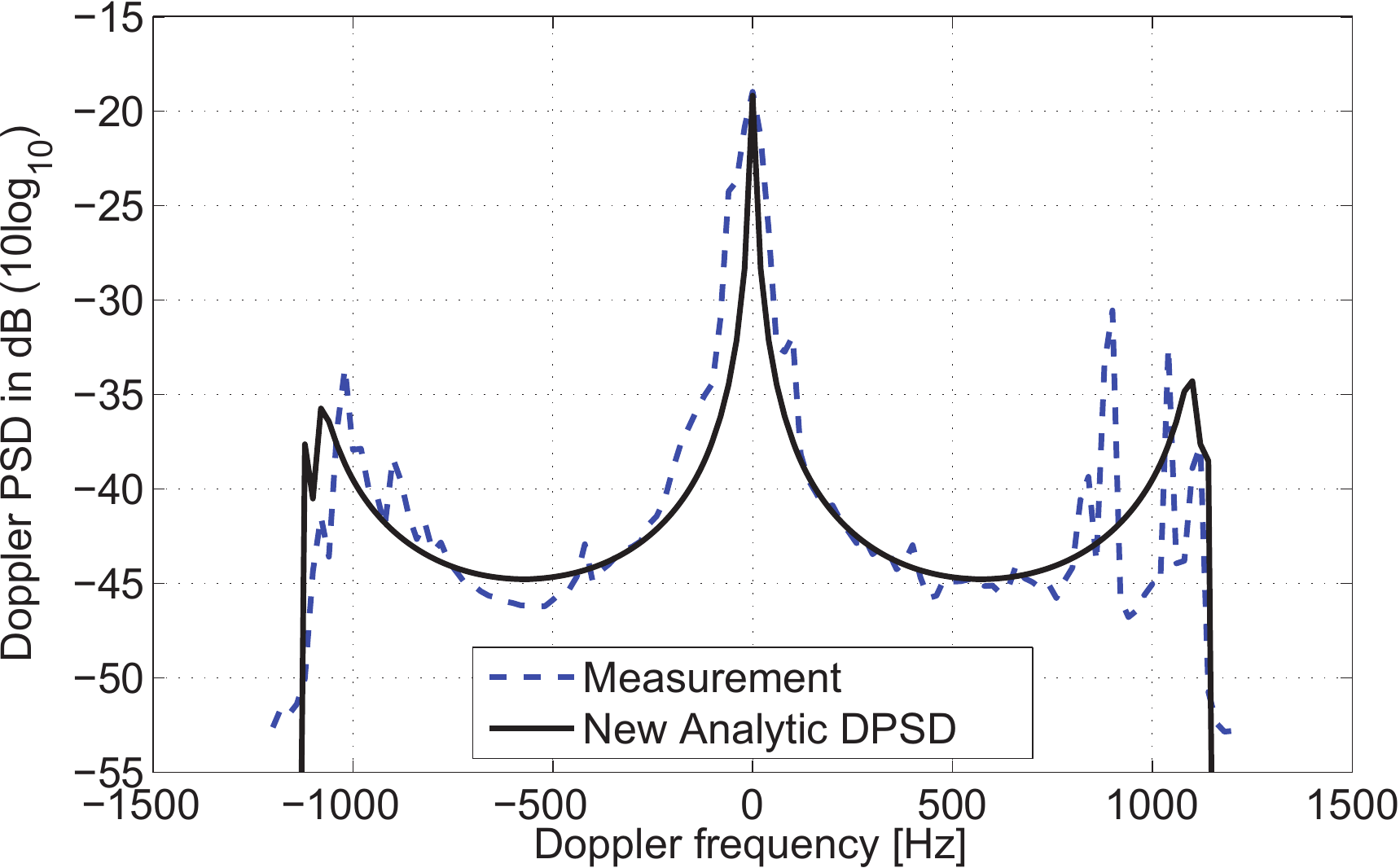}
    \vspace{-0.5cm}
    \caption{A comparison between the new analytic DPSD $S_{hh}\left(\nu\right)$ in (\ref{Proposed_PSD}) and the total measured DPSD of the data set ``MTM-Expressway Same Direction With Wall, 300-400m'' in \cite{Aco07_2}.}
    \label{MComp2}
\vspace{-0.5cm}
\end{figure}

 Fig. \ref{MComp2} shows a comparison between the analytic DPSD $S_{hh}(\nu)$ in (\ref{Proposed_PSD}) and the total measured DPSD, $\tilde{S}^{SD}[m]$. The LoS component of $S_{hh}(\nu)$ is a Dirac delta function, hence excluded in the result for clarity. Before running the optimization, the model and optimization parameters were chosen based on the measurement set up in \cite{Aco07_2} and the optimization performance considerations. Those parameters, together with the local minimum ${\bf x}^{*}$ found after the optimization and the corresponding errors, are summarized in Tables \ref{table1} and \ref{table2}. 
 

The result in Fig. \ref{MComp2} shows that the RSS component of the analytic DPSD is closely matched to the {\it incomplete W-shape} of the total measured DPSD (SD). The error performances in Table \ref{table2} also supports this observation in both numerical and statistical senses, and therefore validating the usefulness of the RSS model. Note that $K=1.535$ was estimated, and this implies that about $40\%$ of the random spectral power is due to signal scattering by RSSs. The rest $60\%$ power, which could not be explained by the RSS part, is mainly concentrated within $|\nu|\le 300$Hz. 
In practice, such power contributions likely come from cars moving in the same direction w.r.t. the Tx and Rx at similar velocities \cite{Zaj14}. Hence, more precise analytic characterization of V2V channels will require the modeling of moving scatterers, such as in 
\cite{Kar09, Zaj14, Yoo16_2}.

\subsection{Model comparison with the data set, ``MTM-Urban Canyon Oncoming 100m.''}

\begin{figure}[t]
    \centering
    \includegraphics [width=8.7cm] {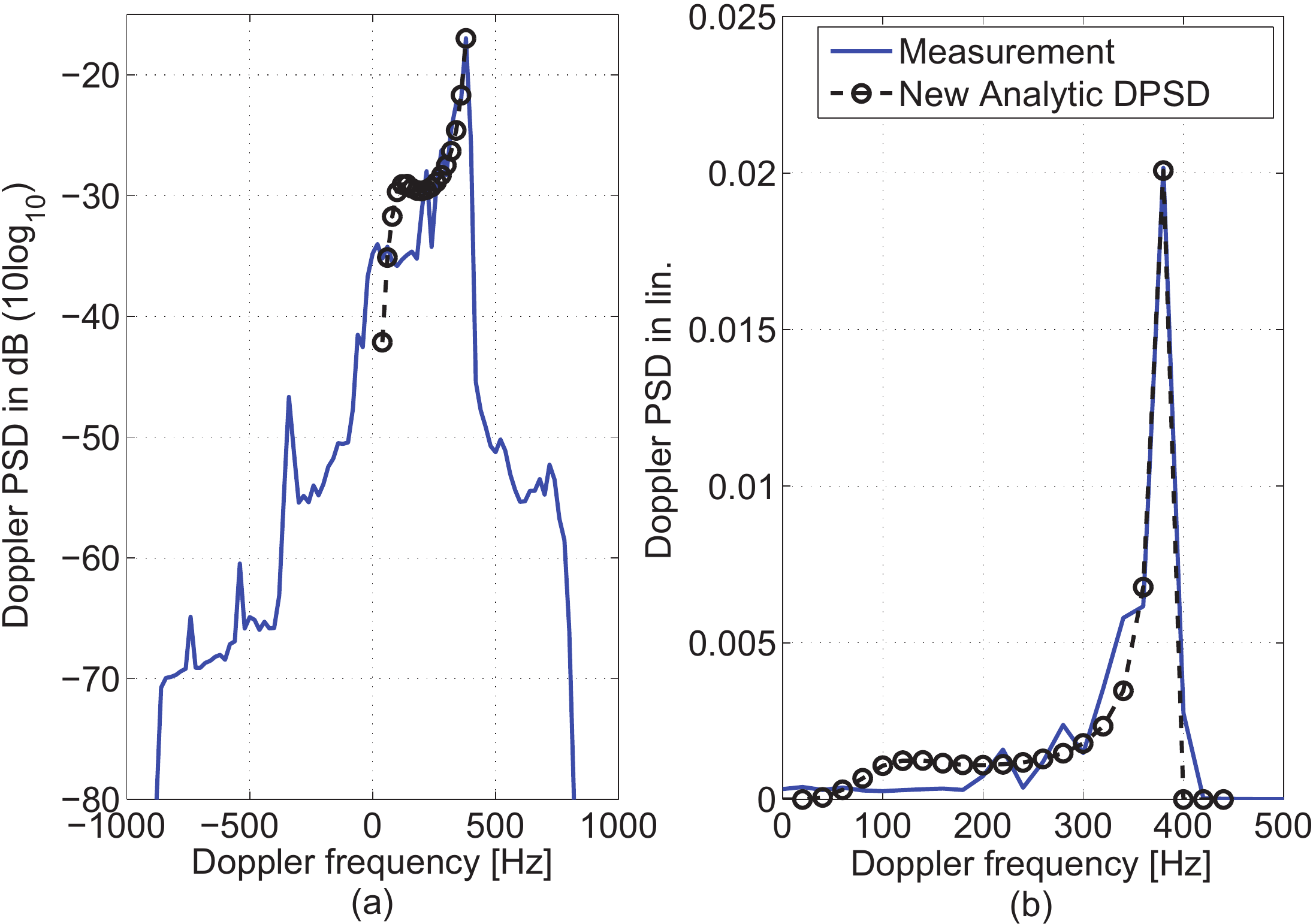}
\vspace{-0.5cm}
    \caption{Comparison results between the new analytic DPSD $S_{hh}\left(\nu\right)$ in (\ref{Proposed_PSD}) and the total measured DPSD of the data set ``MTM-Urban Canyon Oncoming, 100m,'' in \cite{Aco07_2}. (a) dB scale plot; (b) linear scale plot.}
    \label{MComp3}
\vspace{-0.5cm}
\end{figure}

Fig. \ref{MComp3} shows a comparison between the new analytic DPSD, $S_{hh}(\nu)$ in (\ref{Proposed_PSD}), and the total measured DPSD, $\tilde{S}^{OD}[m]$. The model parameters, optimization parameters, and the local minimum ${\bf x}^{*}$ found are summarized in Tables \ref{table1} and \ref{table2}. 
%
The result in Fig. \ref{MComp3} shows that the analytic DPSD is closely matched to the {\it incomplete U-shape} of the measured spectrum (OD) in both dB and linear scales for the Doppler frequency interval, $0 < \nu < f_{D_{\max}}\approx396$Hz over which RSSs generate. The error performances given in Table \ref{table2} also supports this observation. Note that the spectral power outside the range is about $7\%$ of the total PSD power and is likely contributed by moving scatterers. 

In contrast to the SD case, $K=0$ was estimated. This result suggests that most of the received spectral power is contributed from RSSs (such as building surfaces in Fig. 93 of \cite{Aco07_2}) in the street canyon. Such high power contribution can possibly be due to the street canyon effect as pointed out in \cite{Mau04}. Yet, this result should not be exaggerated, as moving scatterers (vehicles) can produce Doppler shifts within $0 < \nu < f_{D_{\max}}\approx396$Hz, over which RSSs generate. Readers may be curious about the deviation between two spectra around $0\le \nu \le 180$Hz and short lengths of the estimated RSS regions. This is primarily due to the EPG assumption in (\ref{eq:RSS_gain}). In OD scenarios, the RSSs closer to the vertices $v_r$ for $r\in\{1,4,5,8\}$ (see Fig. \ref{Subspace}) produce lower Doppler frequencies than $S_n$ close to the midpoint between the Tx and Rx. Those RSSs typically have larger total propagation distances, and hence, considering a proper path-loss exponent (PLE) will reduce the spectral power in that frequency range and also will increase $l_1$ and $l_2$ values. Analytic DPSD solutions of the RSS model, considering a PLE, are not available in the literature, and hence, are definitely worth to investigate in future studies.

\section{Summary and Conclusions}

\FBB{In this paper, an indirect method has been proposed for the DPSD analysis of a generic 2D RSS model for V2V channels. Compared to conventional analytic approaches based on the direct \cite{Ava12} and indirect methods \cite{Wal14, Ava11}, yielding complex multiple integral-form solutions, our method produces a single integral-form DPSD, which is simpler and easier to calculate in computational terms. Our indirect method is based on the Hoeher's theorem and the exact TRV analysis. Hence, the new DPSD solution does not rely on the AoD-AoA independence, assumed in \cite{Ava11} nor requires analytic delay PDF as in \cite{Wal14}. Due to these aspects, our solution is more practical, accurate, and useful for the investigation of the DPSD characteristics due to RSSs, model validation (model parameter estimation) using measurement data, and efficient fading simulator design.}

\FBB{Our DPSD analysis has shown} that transmitted signals spatially spread by RSSs, but partially concentrated in specific joint AoD and AoA angles. This bi-azimuth spread characteristics lead to unique {\it ``incomplete W-shape''} and {\it ``incomplete U-shape''} spectra for SD and OD scenarios, respectively. In the SD scenario, most of the received power was concentrated around $0$ Hz, despite of the large Doppler spread, and even without LoS components. In the OD scenario, the received power was condensed around the maximum Doppler frequency. From numerical analysis, we have found that the length of the RSS region and road width have critical impacts, not only on the shape and Doppler spread of the spectrum, but also its MDS and RDS. The geographical limits of the RSS regions in length, and wide road width can make the channel Doppler spread narrower than the one predicted by the conventional models in \cite{Akk86, Pat05, Zaj08, Che09_2, Zaj14, Zaj15, XChe13, Zaj09, Yua14, Che13}. This {\it spectral shrinkage}, observed in the measured data of \cite{Che13}, was well captured by the finite 2D geometry of the RSS model. Finally, close agreements between the 2D RSS model and the two DPSDs measured in expressway SD and urban caynon OD environments \cite{Aco07_2} have been shown. About $40\%$ of the former and the most of the latter spectra are contributed from RSSs, indicating the importance of RSSs in V2V channels. 

All in all, the research presented herein provides not only a new mathematical framework for modeling and identifying the role of RSSs on the V2V channel dynamics, but also a complementary tool for the channel parameter estimations and fading simulator design based on 2D RSS models. Thus, we believe that the contributions presented in this research will have a significant impact on current V2X communication frameworks, e.g. IEEE and 3GPP standardization circles, 802.11p and 5G for automotive, respectively. Analyzing the DPSD considering PLE and moving scatterers, and its extension to non-stationary channels for time-varying RSS layouts are interesting works in our roadmap.



\section*{Acknowledgment}
The authors gratefully acknowledge the support from Electronic Warfare Research Center at Gwangju Institute of Science and Technology (GIST), originally funded by Defense Acquisition Program Administration (DAPA) and Agency for Defense Development (ADD). This work was also supported by Academy of Finland (grant No. 287249).
\vspace{-0.2cm}

\footnotesize{
}

\end{document}